\documentclass[a4paper,11pt]{article}
\pdfoutput=1

\usepackage{jheppub}

\usepackage[T1]{fontenc} 
\usepackage{subcaption}
\usepackage{slashed}
\usepackage{physics}
\usepackage{wrapfig}
\usepackage{multirow}
\usepackage{siunitx}
\usepackage{booktabs}
\usepackage{anyfontsize}
\usepackage{xcolor}
\usepackage{soul}
\usepackage[tickmarkheight=0.1cm,colorinlistoftodos]{todonotes}

\title{Steady electric currents in magnetized QCD and their use for the equation of state}

\definecolor{ForestGreen}{HTML}{009B55}

\author[a]{B. B. Brandt,}
\author[a]{G. Endr\H{o}di,}
\author[a]{G. Mark\'o,}
\author[a]{A. D. M. Valois}

\affiliation[a]{Universität Bielefeld,  Universitätsstraße 25, 33615 Bielefeld, Germany}

\emailAdd{brandt@physik.uni-bielefeld.de}
\emailAdd{endrodi@physik.uni-bielefeld.de}
\emailAdd{gmarko@physik.uni-bielefeld.de}
\emailAdd{dvalois@physik.uni-bielefeld.de}

\newcommand{\ave}[1]{\left\langle\hspace{0.1cm}#1\hspace{0.1cm}\right\rangle}
\renewcommand{\vec}[1]{\textbf{#1}}
\newcommand{\Z}{\mathcal{Z}}
\newcommand{\val}{\mathrm{val}}
\newcommand{\sea}{\mathrm{sea}}

\newcommand{\ddx}{\dd x\hspace{0.1cm}}

\abstract{
In this paper we study the emergence of steady electric currents in QCD as a response to a non-uniform magnetic background using lattice simulations with $2+1$ quark flavors at the physical point, as well as leading-order chiral perturbation theory. Using these currents, we develop a novel method to determine the leading-order coefficient of the equation of state in a magnetic field expansion: the magnetic susceptibility of the QCD medium. We decompose the current expectation value into valence- and sea-quark contributions and demonstrate that the dominant contribution to the electric current is captured by the valence term alone, allowing for a comparably cheap determination of the susceptibility. Our continuum extrapolated lattice results for the equation of state confirm the findings of some of the existing studies in the literature, namely that the QCD medium behaves diamagnetically at low and paramagnetically at high temperatures.}

\begin{document} 
\maketitle
\flushbottom

\section{Introduction}\label{sec:intro}

Background electromagnetic fields are featured by a series of physical systems ranging from magnetized neutron stars through heavy-ion collisions to the evolution of the early universe~\cite{Kharzeev:2013jha}.
In particular, magnetic fields play a central role for the phenomenology of non-central heavy-ion collisions (HIC).  While a direct measurement of the magnetic field is elusive in such experiments due to its short lifetime, it can be estimated via event-by-event simulations. These simulations predict strong and highly non-uniform electromagnetic fields for peripheral collisions~\cite{Deng:2012pc}, 
with field strengths ranging from $\sqrt{eB} \sim 0.1$ GeV (at typical collisions at RHIC) up to $\sqrt{eB}\sim0.5$ GeV (at the LHC). Such fields are comparable in magnitude to the characteristic scale of the underlying theory, QCD, and therefore might impact the initial stages of the quark-gluon plasma (QGP) created at the experiment and potentially the subsequent hydrodynamic evolution~\cite{Gursoy:2014aka}.
In fact, recent results by the STAR Collaboration at RHIC observed charge-dependent particle flow signals consistent with Faraday and Coulomb effects due to the electromagnetic fields in peripheral collisions~\cite{STAR:2023jdd}.

On the theoretical side, the impact of electromagnetic fields on QCD matter can be studied by lattice QCD simulations as well as within low-energy QCD models and effective theories (see~\cite{Andersen:2014xxa} for a review on the topic).
So far, most of these studies have assumed spatially homogeneous fields. A theoretical improvement to this picture is to consider a non-uniform magnetic field that more closely resembles the situation in the early stages of off-central collisions.
For a specific, analytically treatable inhomogeneous profile~\cite{Cangemi:1995ee},
a study using the NJL model~\cite{Cao:2017gqs} predicted non-trivial structures appearing in various observables relevant for thermodynamics. Recent lattice QCD simulations revealed that such fields not only induce an inhomogeneous breaking of chiral symmetry, but also affect the confinement aspects of the theory in a non-uniform manner~\cite{Brandt:2023dir}. In addition, the inhomogeneous background fields are expected to generate local electric currents in equilibrium QCD~\cite{Cao:2017gqs}.

In general, conserved currents appear as a manifestation of global symmetries. In non-relativistic quantum mechanical systems in equilibrium, the global flow of conserved currents in the ground state is forbidden by Bloch's theorem~\cite{PhysRev.75.502}. However, the theorem does not forbid a local current density, as long as its spatial integral vanishes. In particular, this is the case for an electric current due to an inhomogeneous background magnetic field in accordance with Amp\`{e}re's law. This naturally leads us to the question: could inhomogeneous magnetic fields induce local currents in an equilibrium state of QCD and if yes, what role do they play for QCD thermodynamics?

In this paper, we perform lattice simulations to investigate the emergence of electric current densities as a response of the strongly interacting medium to a non-uniform magnetic field in an equilibrium setup. 
Our simulations employ 2+1 flavors of stout-improved rooted staggered quarks with physical masses and a continuum extrapolation based on four lattice spacings.
Using the currents measured on the lattice and Amp\`{e}re's law, we present a novel approach to compute the leading-order coefficient of the equation of state in a $B$-expansion, i.e.\, the magnetic susceptibility of the thermal QCD medium.
This new determination of the susceptibility may be compared with existing lattice techniques in the literature~\cite{Levkova:2013qda,Bali:2013esa,Bonati:2013lca,Bali:2014kia,Bali:2020bcn,Buividovich:2021fsa}.
Our results show that, in the continuum limit, the magnetic susceptibility changes sign across the QCD crossover, characterized by the pseudo-critical temperature $T_c$, indicating that the response of the QCD medium to an external magnetic field comes in at least two phases: 1.\ a weak diamagnetic phase at temperatures below $T_c$, and 2.\ a strong paramagnetic phase at temperatures above $T_c$. This confirms the findings of~\cite{Bali:2020bcn}.

In addition to computing the electric current expectation value in an ensemble fully incorporating the inhomogeneous magnetic field background, we will introduce what we call the valence approximation, based on the weak-field expansion of fermionic expectation values into sea and valence contributions (see App.~\ref{app:sea_valence}), introduced for the chiral condensate in Ref.~\cite{DElia:2011koc}. We empirically show that, unlike the chiral condensate, the electric current has a negligible sea contribution across a broad range of temperatures and magnetic fields. As a consequence, the electric current at a given magnetic field strength $B$ is well-approximated by the valence term, which only requires gauge configurations at $B=0$. This observation decreases the computational cost by a substantial factor. 
Thereby, we suggest that our method could be applied to compute the magnetic susceptibility using different fermion discretizations using less computational resources. To the best of our knowledge, the calculation of the magnetic susceptibility has only been done in the staggered formulation so far. 

Our lattice simulations are corroborated by a determination of the electric current induced by the inhomogeneous magnetic field within a low-energy effective theory of QCD, chiral perturbation theory. We work out the details of this approach in App.~\ref{app:chiPT}.

This article is organized as follows. In Sec.~\ref{sec:mag_susc}, we discuss our method to obtain the magnetic susceptibility from electric currents using Amp\`{e}re's law. In Sec.~\ref{sec:setup} we describe our simulation setup and define the electric current density in the staggered formulation. This is followed by  Sec.~\ref{sec:renormalization}, which is devoted to the renormalization of the current operator as well as that of  the susceptibility. In Sec.~\ref{sec:results} we present our lattice results from the full-current and the valence approximation and finally, in Sec.~\ref{sec:conclusions} we conclude.

\section{Magnetic susceptibility and Amp\`{e}re's law}\label{sec:mag_susc}

In classical electrodynamics, the total magnetic field in a magnetized medium is given by
\begin{equation}
\vec{B} = \vec{H}+e\vec{M}\,,
\label{eq:total_B}
\end{equation}
where $\vec{H}$ is the external field that one would apply to the system in an experiment and $\vec{M}$ is the magnetization, i.e.\ the response of the system to the applied field. Therefore, $\vec{B}$ represents the total magnetic field that would be measured by an external observer. 
In Eq.~\eqref{eq:total_B}, we used natural units for the vacuum magnetic permeability $\mu_0 = 1$. Moreover, we expressed the magnetization in units of the elementary electric charge $e>0$ -- this will be convenient later when we discuss renormalization.
To each field in Eq.~\eqref{eq:total_B} there are associated electric current densities, which follow a similar relation
\begin{equation}
\vec{j}_{B} = \vec{j}_{H} + \vec{j}\,,
\label{eq:total_j}
\end{equation}
where $\vec{j}_{H}$ is the free current density that generates $\vec{H}$, $\vec{j}$ is the magnetization current density, and $\vec{j}_{B}$ is the total current density. Since we are interested in the response of the QCD medium to the background field, we will focus on $\vec{j}$. 

In a static medium, the magnetization at weak magnetic fields is related to the total field by
\begin{equation}
\vec{M}(\vec{r}) = \frac{1}{V}\int \dd^3\vec{r}^{\prime}\, \chi(\vec{r}-\vec{r}^{\prime})\,e\vec{B}(\vec{r}^{\prime})\,,
\label{eq:magnetization}
\end{equation}
where $V$ is the volume, $\chi$ the local magnetic susceptibility and we neglected terms of $\mathcal{O}(B^3)$. In general, $\chi$ may be a function of the coordinates, but the translational invariance of the medium constrains this dependence. We made this fact explicit by writing $\chi(\vec{r},\vec{r}^{\prime}) = \chi(\vec{r}-\vec{r}^{\prime})$. The magnetization current can be obtained from $\vec{M}$ via Amp\`{e}re's law for the magnetization,
\begin{equation}
\vec{j}(\vec{r}) = \curl{\vec{M}(\vec{r})} = \frac{1}{V}\int \dd^3\vec{r}^{\prime} \,\chi(\vec{r}^{\prime})\,\curl{e\vec{B}(\vec{r}-\vec{r}^{\prime})}\,,
\label{eq:ampere_law}
\end{equation}
where we integrated by parts and relabeled $\vec{r}^{\prime}\to \vec{r}-\vec{r}^{\prime}$, making use of translation invariance again. 

Next, we specialize to a magnetic field of the form $\vec{B}(\vec{r}) = B(x) \,\vec{e}_z$, where $x$ is the first component of the vector $\vec{r}$. In this case, Eq.~\eqref{eq:ampere_law} implies that the current $\vec{j}$ points in the $y$ direction and is given by
\begin{equation}
j_y(x) = \frac{1}{L}\int \dd x^{\prime}\, \chi(x^{\prime})\,\pdv{eB(x-x^{\prime})}{x}\,,
\label{eq:jy_convolution}
\end{equation}
The Fourier transform of Eq.~\eqref{eq:jy_convolution} gives
\begin{equation}
\widetilde{j}_y(p) = ip\,\widetilde{\chi}(p)\,e\widetilde{B}(p)\,,
\label{eq:chip1}
\end{equation}
where $p$ denotes the momentum in the $x$ direction. This result is still obtained within a weak-field expansion, i.e.\ we neglected $\mathcal{O}(B^3)$ terms on the right hand side. 
In this work, we will be mostly interested in the susceptibility at zero momentum. This encodes the response of the medium to homogeneous magnetic fields, and we will denote the corresponding, leading weak-field coefficient as $\chi_m=\widetilde{\chi}(p=0)$.

Alternatively to the above derivation, the magnetic susceptibility $\chi_m$
can be obtained directly from the partition function $\Z$ of the system in a homogeneous background magnetic field, oriented along the $z$ direction without loss of generality,
\begin{equation}
\chi_m \equiv \left.\frac{T}{V}\frac{\partial^2\log\Z}{\partial(eB)^2}\right|_{B=0}\,,
\label{eq:chi_m_thermo}
\end{equation}
where $T$ denotes the temperature.
This definition of $\chi_m$ is consistent with the $p=0$ limit of $\widetilde{\chi}(p)$ considered in Eq.~\eqref{eq:chip1}. The factors of $e$ appear in a manner so that the derivative of $\log\mathcal{Z}$ in Eq.~\eqref{eq:chi_m_thermo} is taken with respect to the renormalization-group-invariant combination $eB$. 

\subsection{Susceptibility on the torus}

The discussion so far corresponds to the infinite volume system. Next we will consider a finite volume with periodic boundary conditions, in order to comply with the standard setup in lattice simulations. Specifically, we take our volume to be a symmetric box with side length $L$, and label the coordinates so that $-L/2\le x \le L/2$.
Furthermore, we will work with an inhomogeneous magnetic field profile,
\begin{equation}
B(x) = B\cosh^{-2}\qty(\frac{x}{\epsilon})\,,
\label{eq:inv_cosh_profile}
\end{equation}
where $\epsilon$ is the inhomogeneity parameter. In the discussion below we do not exploit the specific form of $B(x)$, but only the fact that it is an even function, $B(-x)=B(x)$. The latter also implies $j_y(-x)=-j_y(x)$.

Let us now come back to Eq.~\eqref{eq:chip1}, and focus on the zero-momentum limit of $\widetilde{\chi}(p)$. In the infinite volume, when $p$ is continuous, this can be achieved via L'H\^{o}pital's theorem,
\begin{equation}
\widetilde{\chi}(p=0) = -i\lim_{p\to0}\frac{\widetilde{j}_y(p)}{p\,e\widetilde{B}(p)} = -i\lim_{p\to0}\frac{\dfrac{\partial\widetilde{j}_y(p)}{\partial p}}{e\widetilde{B}(p) + p\dfrac{\partial e\widetilde{B}(p)}{\partial p}} = -\frac{1}{e\widetilde{B}(0)}\int \ddx x \,j_y(x)\,.
\label{eq:susc_p=0_limit}
\end{equation}
On the torus, however, $p$ is discrete, and we are formally not allowed to take the $p\to0$ limit. Therefore, we consider $\widetilde{\chi}(p)$ at finite momentum and rewrite it using the parity symmetry of the current, $j_y(-x) = -j_y(x)$, and the magnetic field, $B(-x) = B(x)$,
\begin{equation}
\widetilde{\chi}(p) = -i\frac{\widetilde{j}_y(p)}{p\, e\widetilde{B}(p)} = -i\dfrac{\int_{-L/2}^{L/2} \ddx e^{-ipx} j_y(x)}{p\int_{-L/2}^{L/2} \ddx e^{-ipx} eB(x)} = \dfrac{\int_{0}^{L/2} \ddx \dfrac{\sin(px)}{p} j_y(x)}{\int_{0}^{L/2} \ddx \dfrac{\sin(px)}{p} \dfrac{\partial eB(x)}{\partial x}}\,, \label{eq:chi_p_space}
\end{equation}
where in the last equality we integrated by parts. For low $p$, we can approximate the sine in the integral above using its Taylor series. For small $x$, we can expand the sine around 0, whereas for $x\sim L/2$, we can expand it around $L/2$. Therefore, in the $p\to0$ limit, we approximate the sine in Eq.~\eqref{eq:chi_p_space} by a wedge-like function in the interval $[0,L/2]$,
\begin{align}
W(x,L) \equiv \left \{
    \begin{array}{ll}
        x  &\, \mbox{for } 0 \leq x \leq \frac{L}{4}\,, \\
        \dfrac{L}{2}-x &\, \mbox{for } \frac{L}{4} \leq x \leq \frac{L}{2}\,,
    \end{array}\right. 
    \label{eq:wedge}
\end{align}
which has the same parity symmetry as the sine in that interval, and it satisfies the boundary conditions.

Hence, the zero-momentum susceptibility becomes
\begin{equation}
\widetilde{\chi}(p=0) = -\dfrac{\int_{0}^{L/2} \ddx W(x,L)\,j_y(x)}{\int_{0}^{L/2} \ddx W(x,L) \dfrac{\partial eB(x)}{\partial x}}\,. \label{eq:current_convol_with_wedge}
\end{equation}
Notice that the magnetic field~\eqref{eq:inv_cosh_profile} decays exponentially towards the boundaries of the volume and so does the electric current. Therefore, the contribution from $0 \leq x \leq L/4$ dominates the integral as $L\to\infty$ and Eq.~\eqref{eq:current_convol_with_wedge} approaches Eq.~\eqref{eq:susc_p=0_limit} in the infinite-volume limit. In Sec.~\ref{sec:results}, we will apply Eq.~\eqref{eq:current_convol_with_wedge} to calculate $\widetilde{\chi}(0)$ on the lattice. 

Just like Eq.~\eqref{eq:chip1}, all equations above are formulated as a weak-field expansion, with $\mathcal{O}(B^3)$ terms neglected.
To recover the magnetic susceptibility~\eqref{eq:chi_m_thermo} from the observables determined at $B>0$, we need to extrapolate to $B=0$,
\begin{equation}
\chi_m = -\lim_{B\to0}\dfrac{\int_{0}^{L/2} \ddx W(x,L)\,j_y(x)}{\int_{0}^{L/2} \ddx W(x,L) \dfrac{\partial eB(x)}{\partial x}} \,. \label{eq:current_convol_with_wedge_B0_limit}
\end{equation}
At $B=0$, both $j_y(x)$ and $B(x)$ vanish but the ratio in the right-hand side of Eq.~\eqref{eq:current_convol_with_wedge_B0_limit} remains finite. Furthermore, as we will discuss in Sec.~\ref{sec:setup}, in a finite box the magnetic field is quantized and we only have access to discrete values of $B$. Therefore we extrapolate to vanishing fields using the ansatz $c_1 + c_2(eB)^2$, valid for low $B$, based on the symmetry of the partition function $\Z(B) = \Z(-B)$, which only allows for even powers of $B$ in the susceptibility. We obtained the parameters $c_1$ and $c_2$ by fitting our finite $B$ lattice data.

The so obtained $\chi_m$ contains an additive logarithmic divergence which has to be cancelled via renormalization. We will discuss its origin and our renormalization scheme for $\chi_m$ and $j_y$ in Sec.~\ref{sec:renormalization}.

\section{Simulation setup}\label{sec:setup}
\subsection{Non-uniform magnetic fields on the lattice}\label{sec:B_on_the_lattice}
In a finite box, the flux of the magnetic field is quantized. In order to couple a quark flavor $f$, with charge $q_f$, to a magnetic field background given by Eq.~\eqref{eq:inv_cosh_profile} with amplitude $B$ in the $z$-direction, where $-L_x/2 \leq x < L_x/2$, we need the following U(1) factors (see \cite{Brandt:2023dir} for more details),
\begin{align}
u^f_{x}(x,y,z,t) &= 
    \left\{
        \begin{array}{ll}
        e^{-i2\pi N_b (y/L_y + 1/2)} \qquad & \mbox{if } x = L_x/2-a\,, \nonumber \\
        1 & \mbox{otherwise}\,,
        \end{array}
    \right. \\
u^f_{y}(x,y,z,t) &= e^{iq_fB\epsilon a\qty[\tanh(x/\epsilon) + \tanh(L_x/2\epsilon)]}\,, \label{eq:links1} \\
u^f_z(x,y,z,t) &= 1\,, \nonumber\\
u^f_t(x,y,z,t) &= 1\,. \nonumber
\end{align}
in addition to a quantization condition for the magnetic flux in a finite box,
\begin{equation}
qB = \frac{\pi N_b}{L_y\epsilon\tanh(L_x/2\epsilon)},\hspace{0.5cm} N_b\in\mathbb{Z}.
\label{eq:quantization}
\end{equation}
Here, $q$ is a reference charge. For a many-flavor system, the reference charge is given by the modulus of the greatest common divisor of the charges. In the case of $u$, $d$, and $s$ quarks: $q = |q_d| = |q_s|= e/3$. The parameter $\epsilon$ controls the width of the profile. The value of $\epsilon$ was chosen such that $B(-L_x/2) = B(L_x/2) \approx 0$ to avoid substantial effects from the boundaries.

\begin{figure}[!h]
    \centering
    \includegraphics{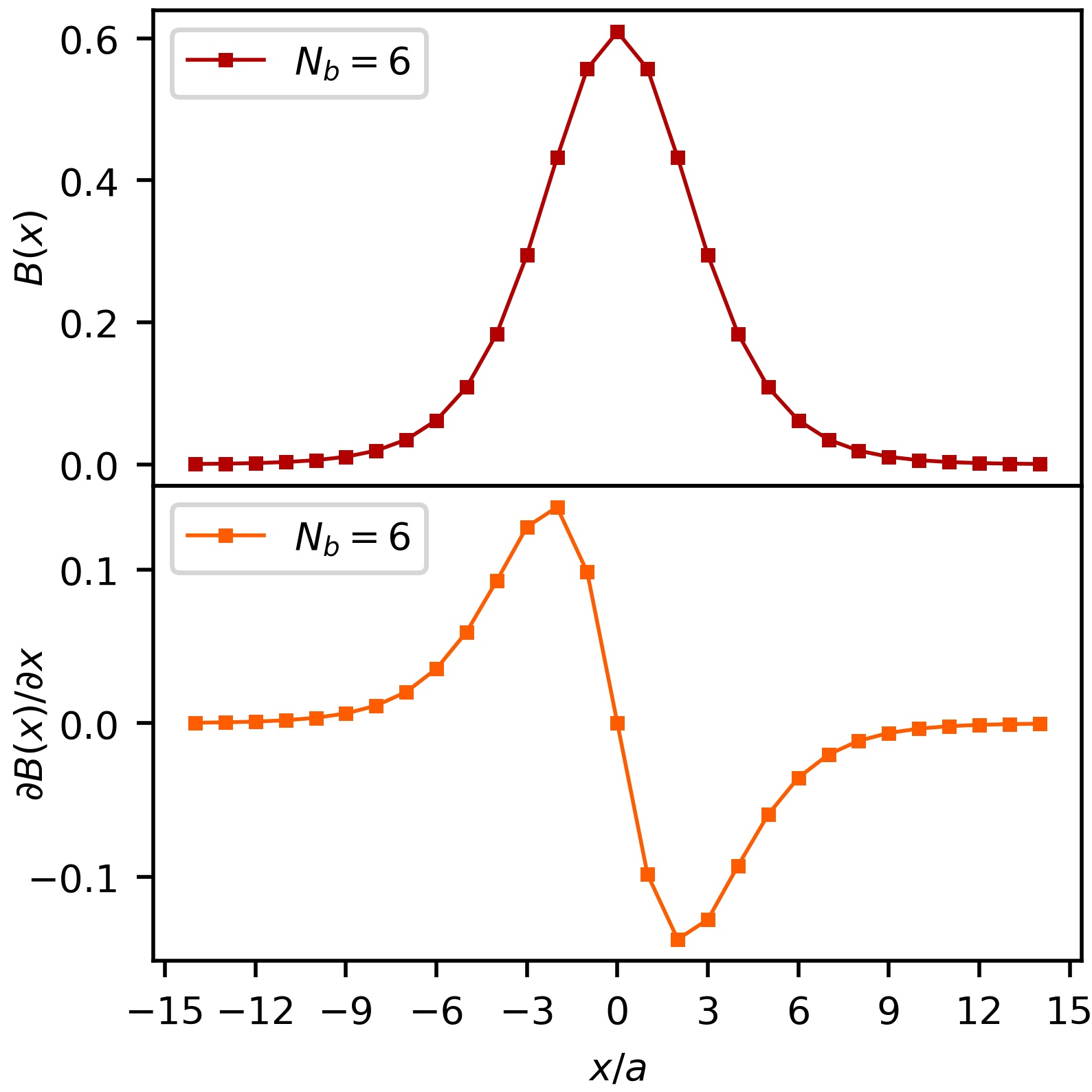}
    \caption{\small Magnetic field profile (upper plot) and its first derivative (lower plot) in lattice units for $N_b = 6$ as a function of the $x$ coordinate. The connecting line are only included to guide the eye.}
    \label{fig:B_current_profile}
\end{figure}

Rewriting Eq.~\eqref{eq:quantization} in terms of the aspect ratio $N_s/N_t$, where $N_s$ and $N_t$ are the number spatial and temporal sites, respectively
\begin{equation}
qB = \pi N_b T \qty[\frac{N_s}{N_t}\epsilon \tanh(\dfrac{N_s}{N_t}\dfrac{1}{2\epsilon T})]^{-1}\,.
\label{eq:quantization_aspec_rat}
\end{equation}
where $T = 1/aN_t$. This equation implies that at fixed $T$, $\epsilon$, and $N_b$ the physical value of the magnetic field $B$ is the same between different ensembles if they have the same aspect ratio. This observation will be useful in Sec.~\ref{sec:full} when we discuss the results. In Fig.~\ref{fig:B_current_profile} we illustrate the magnetic field profile implemented by the links in \eqref{eq:links1} and its first spatial derivative, which will be relevant for the following discussion regarding the electric current. 
\subsection{Full-current setup}\label{sec:electric_current}
We simulated $2+1$ flavors of rooted staggered fermions with physical masses. The Dirac operator $\slashed{D}$ is a one-link operator including two steps of stout smearing. 
In the rooted staggered formalism, the $\mu$-component of the electric current density is given by
\begin{equation}
\ave{\bar{\psi}\Gamma_{\mu}\psi_f(x)}_{B,T} = \frac{T}{4L_yL_z \mathcal{Z}(B)}\int \mathcal{D}U\hspace{0.1cm}e^{-S_g}\Tr_x[(\Gamma_{\mu}\mathcal{M}^{-1})(q_fB)]\prod_{f^{\prime}}\det\mathcal{M}(q_{f^{\prime}}B)^{1/4}\,, \label{eq:staggered_current}
\end{equation}
where $\mathcal{M}(q_fB) = \slashed{D}(U,q_{f}B)+m_{f}$ is the massive Dirac operator, and $\Tr_x$ represents the trace over $y$, $z$, $t$, and color but keeping the $x$ coordinate fixed. $\Gamma_{\mu}$ are the staggered equivalents of the Euclidean Dirac matrices, defined as in~\cite{Durr:2013gp},
\begin{equation}
\Gamma_{\mu}(\vec{x},\vec{x}^{\prime}) = \frac{1}{2}\eta_{\mu}(\vec{x})\left[U_{\mu}(\vec{x})u_{\mu}^f(\vec{x})\delta_{\vec{x}^{\prime},\vec{x}+\hat{\mu}} + U_{\mu}^{\dagger}(\vec{x}-\hat{\mu})u_{\mu }^{f*}(\vec{x}-\hat{\mu})\delta_{\vec{x}^{\prime},\vec{x}-\hat{\mu}}\right]\,,
\label{eq:Gammamu_def}
\end{equation}
where $\eta_{\mu}(\vec{x})$ are the staggered phases, and $u_{\mu}^f(\vec{x})$ are the U(1) gauge links, defined in Eq.~\eqref{eq:links1}. To avoid confusion with the $x$ coordinate, we denoted the spacetime dependence of the $\Gamma$-matrices as $\vec{x} \equiv (x,y,z,t)$. Note that in Eq.~\eqref{eq:Gammamu_def} we made the dependence of $\Gamma_\mu$ on the quark flavor implicit. The total electric current density due to all quark flavors is given by
\begin{equation}
j_{\mu}(x,T,B) = \sum_f \frac{q_f}{e} \ave{\bar{\psi}\Gamma_{\mu}\psi_f(x)}_{B,T}\,.
\label{eq:all_flavor_curr}
\end{equation}
Note that this definition of the current is consistent with Eq.~\eqref{eq:ampere_law}.

For our magnetic field profile, the only non-zero component is for $\mu=y$. In Sec.~\ref{sec:valence_fixed_eps_physical}, we will also discuss the correlator between the electric current and the Polyakov loop and between the chiral condensate and the Polyakov loop. These operators can be defined in a similar fashion \cite{Brandt:2023dir}
\begin{align}
\ave{\bar{\psi}\psi_f(x)}_{B,T} &= \frac{T}{4L_yL_z}\frac{1}{\mathcal{Z}(B)}\int \mathcal{D}U\hspace{0.1cm} e^{-S_g}\Tr_x\left[\mathcal{M}^{-1}(q_fB)\right]\prod_{f'}\left[\det\mathcal{M}(q_{f^{\prime}}B)\right]^{1/4}\,, \label{eq:local_chiral_condensate}\\
\ave{P(x)}_{B,T} &= \frac{1}{L_yL_z}\frac{1}{\mathcal{Z}(B)}\int \mathcal{D}U\hspace{0.1cm} e^{-S_g}\prod_{t}\sum_{y,z}\Re\Tr_x[U_t(x,y,z,t)]\prod_{f}\left[\det\mathcal{M}(q_fB)\right]^{1/4}\,.\label{eq:local_pol}
\end{align} 
In this setup, we computed the electric current operator in Eq.~\eqref{eq:all_flavor_curr} using four different lattice spacings in order to study its continuum limit. To resolve the $x$-dependence in the fermionic traces, we employed random sources in the same way as in Ref.~\cite{Brandt:2023dir}. Our parameter set covers a range of temperatures from $113 \text{ MeV}$ to $300 \text{ MeV}$, and various values for the magnetic flux quantum $N_b$. Moreover, we fixed the physical value of the inhomogeneity parameter to $\epsilon \approx 0.6$ fm, which is compatible with what one would expect from model simulations of off-central heavy-ion collisions~\cite{Deng:2012pc}. For the renormalization of some observables, we also used $T\approx0$ gauge configurations. In Table~\ref{tab:full_setup}, we show the simulation parameters used in the full-current setup.
\begin{table}[!h]
\fontsize{10pt}{10pt}\selectfont
\centering
  \begin{tabular}{c|c|p{0.03\textwidth}p{0.03\textwidth}p{0.03\textwidth}p{0.03\textwidth}p{0.03\textwidth}p{0.03\textwidth}p{0.03\textwidth}p{0.03\textwidth}p{0.03\textwidth}p{0.03\textwidth}p{0.03\textwidth}p{0.03\textwidth}}
    \toprule
    \multicolumn{13}{c}{$T>0$ ensemble} \\
    \midrule
    \multirow{2}{*}{$24^3\times 6$} & $T$ (MeV) & 187 & 217 & 248 & 280 & 300 \\
      & $\epsilon/a$ & 3.29 & 3.83 & 4.36 & 4.94 & 5.28\\
      \midrule
      \multirow{2}{*}{$24^3\times 8$} & $T$ (MeV) & 68 & 113 & 124 & 142 & 148 & 155 & 162 & 176 & 187 & 250 & 280 & 300 \\
      & $\epsilon/a$ & 1.60 & 2.66 & 2.92 & 3.34 & 3.48 & 3.65 & 3.81 & 4.13 & 4.40 & 5.88 & 6.58 & 7.04 \\
      \midrule
      \multirow{2}{*}{$28^3\times 10$} & $T$ (MeV) & 113 & 124 & 142 & 148 & 155 & 162 & 176 & 200 & 250 & 300 \\
      & $\epsilon/a$ & 3.32 & 3.65 & 4.17 & 4.34 & 4.56 & 4.75 & 5.17 & 5.88 & 7.33 & 8.81\\
      \midrule
      \multirow{2}{*}{$36^3\times 12$} & $T$ (MeV) & 113 & 124 & 142 & 148 & 155 & 162 \\
      & $\epsilon/a$ & 3.98 & 4.38 & 5.00 & 5.22 & 5.46 & 5.71\\
      \midrule
      \multicolumn{13}{c}{$T\sim0$ ensemble} \\
      \midrule
      \multirow{2}{*}{$24^3\times 32$} & $T$ (MeV) & 13 & 21 & 29\\
      & $\epsilon/a$ & 1.20 & 2.00 & 2.73 \\  
    \bottomrule
  \end{tabular}
  \caption{\small Simulation parameters for the full-current setup. To keep the physical value of the magnetic profile width fixed at $\epsilon\approx$ 0.6 fm, we tuned the ratio $\epsilon/a$ for every temperature.}
  \label{tab:full_setup}
\end{table}
\subsection{Valence-current setup}\label{sec:valence}
Above we have seen that the magnetic field appears in the current expectation value~\eqref{eq:staggered_current} in two different ways: in the current operator (valence contribution) as well as in the fermion determinant (sea contribution). In Sec.~\ref{sec:valence_fixed_eps_physical} we will demonstrate that the former contribution is by far the dominant one and therefore it is a good approximation to only include the magnetic field in the operator.
For this valence approximation, we also employed rooted staggered fermions at the physical point with two stout smearing steps. We start by decomposing the same staggered current into operators that are relevant to quantify the impact of sea and valence effects in the weak-field regime,
\begin{align}
\ave{\bar{\psi}\Gamma_{\mu}\psi_f(x)}_{B,T}^{\val} &= \frac{T}{4L_yL_z \mathcal{Z}(0)}\int \mathcal{D}U\hspace{0.1cm}e^{-S_g}\Tr_x[(\Gamma_{\mu}\mathcal{M}^{-1})(q_fB)]\prod_{f^{\prime}}\det\mathcal{M}(0)^{1/4} , \label{eq:val_term}\\
\ave{\bar{\psi}\Gamma_{\mu}\psi_f(x)}_{B,T}^{\sea} &= \frac{T}{4L_yL_z \mathcal{Z}(B)}\int \mathcal{D}U\hspace{0.1cm}e^{-S_g}\Tr_x[(\Gamma_{\mu}\mathcal{M}^{-1})(0)]\prod_{f^{\prime}}\det\mathcal{M}(q_{f^{\prime}}B)^{1/4} , \label{eq:sea_term}
\end{align}
where the superscripts indicate the valence and sea terms, respectively. We can also define the valence and sea currents due to all flavors as 
\begin{align}
j^{\val}_{\mu}(x,T,B) & = \sum_f \frac{q_f}{e} \ave{\bar{\psi}\Gamma_{\mu}\psi_f(x)}_{T,B}^{\val} \quad \text{and} \\
j^{\sea}_{\mu}(x,T,B) & = \sum_f \frac{q_f}{e} \ave{\bar{\psi}\Gamma_{\mu}\psi_f(x)}_{T,B}^{\sea}\,,
\end{align}
respectively. For a weak magnetic field, we can show that these currents follow an additivity relation (see Appendix \ref{app:sea_valence} for more details)
\begin{equation}
j_{\mu} = j^{\val}_{\mu} + j^{\sea}_{\mu} + \mathcal{O}(B^3) \,.
\label{eq:sea_valence_J}
\end{equation}
From now on, we refer to the terms on the right-hand side as valence and sea currents.

On the lattice, the sea current corresponds to a $B$-independent operator computed on configurations at $B\neq0$, whereas the valence one corresponds to a $B$-dependent operator computed on gauge configurations at $B=0$. This makes the valence term computationally much cheaper than the sea, since we only need a single ensemble at $B=0$ to resolve the $B$-dependence of $j_{\mu}^{\val}$.

Notice that both currents contain all orders in $B$. The $\mathcal{O}(B^3)$ notation in Eq.~\eqref{eq:sea_valence_J} simply emphasizes that the decomposition of $j_{\mu}$ is only exact at leading order in the magnetic field. In Sec.~\ref{sec:valence_fixed_eps_physical}, we empirically show that $j_{\mu}^{\sea}$ is negligible for a broad range of temperatures and up to rather strong magnetic fields. This result justifies what we call the valence approximation: $j_{\mu} \approx j_{\mu}^{\val}$, where the dominant contribution to the full current is captured by the valence term.

Within the valence approximation, we explored two cases:

\begin{itemize}
    \item \textbf{Constant $\epsilon$:} to validate the valence approximation, we computed the valence current using the same set of parameters from Table \ref{tab:full_setup}, i.e.\ setting $\epsilon \approx 0.6$ fm, and compared it to the full current. We will discuss the corresponding results in Sec.~\ref{sec:valence_fixed_eps_physical}.
    
    \item \textbf{Constant $\epsilon/L$:}  to have a better control over finite volume effects at hight $T$, we computed the valence current using the parameters from Table~\ref{tab:valence_setup2}, i.e.\ with fixed $\epsilon/L = 0.12$ for all $T$ and lattices. This ensures that the inhomogeneity of the magnetic field fits into the box regardless of the parameters. Note that the susceptibility defined via Eq.~\eqref{eq:current_convol_with_wedge_B0_limit} does not depend on $\epsilon/L$ in the limit of $B\to0$. In this case, we used existing ensembles from~\cite{Bali:2020bcn}. We will discuss the corresponding results in Sec.~\ref{sec:valence_2}.
\end{itemize}

 The reason for studying two cases was to understand whether boundary effects become important at high $T$. Since the physical value of $\epsilon$ is fixed in the first case of the valence setup, as well as in the full-current setup, the ratio $\epsilon/L$ increases at high $T$, where due to the smaller lattice spacings, the physical volume decreases. This could mean large contributions to the susceptibility coming from the edges. Therefore, the second case of the valence setup was designed to check if the high-$T$ part of the susceptibility is affected by the boundary effects.
 
At this point, we make two important remarks on the computation of the susceptibility in the valence setup. First, using the sea-valence expansion in Eq.~\eqref{eq:current_convol_with_wedge} yields
\begin{equation}
\widetilde{\chi}(p=0) = -\dfrac{\int_{0}^{L/2} \ddx W(x,L)j^{\val}_y(x)}{\int_{0}^{L/2} \ddx W(x,L) \dfrac{\partial eB(x)}{\partial x}}-\dfrac{\int_{0}^{L/2} \ddx W(x,L)j^{\sea}_y(x)}{\int_{0}^{L/2} \ddx W(x,L) \dfrac{\partial eB(x)}{\partial x}} + \mathcal{O}(B^2)\,.
 \end{equation}
Therefore, the decomposition of the susceptibility is also of the form $\chi_m=\chi_m^{\val}+\chi_m^{\sea}$. In other words, the accuracy of using the valence approximation to compute the susceptibility is only limited by the magnitude of the sea term and is independent of the higher-order terms, for they vanish in the $B=0$ limit.
Second, since the valence currents for different values of $B$ are computed on the same gauge configurations, we employed a correlated fit when carrying out the $B=0$ extrapolation to account for the correlation of the data.
\begin{table}[!h]
\fontsize{10pt}{10pt}\selectfont
\centering
\setlength{\tabcolsep}{5pt}
  \begin{tabular}{c|c|p{0.03\textwidth}p{0.03\textwidth}p{0.03\textwidth}p{0.03\textwidth}p{0.03\textwidth}p{0.03\textwidth}p{0.03\textwidth}p{0.03\textwidth}p{0.03\textwidth}p{0.03\textwidth}p{0.03\textwidth}p{0.03\textwidth}p{0.03\textwidth}p{0.03\textwidth}}
    \toprule
    \multicolumn{13}{c}{$T>0$ ensemble} \\
    \midrule
    \multirow{2}{*}{$24^3\times 6$} & $T$ (MeV) & 189 & 203 & 220 & 264 & 305 \\
      & $\epsilon/a$ & \multicolumn{11}{c}{2.88}\\
      \midrule
      \multirow{2}{*}{$24^3\times 8$} & $T$ (MeV) & 123 & 137 & 142 & 147 & 163 & 174 & 186 & 198 & 210 & 250 & 280 & 303\\
      & $\epsilon/a$ & \multicolumn{11}{c}{2.88} \\
      \midrule
      \multirow{2}{*}{$28^3\times 10$} & $T$ (MeV) & 106 & 114 & 123 & 130 & 135 & 139 & 144 & 154 & 158 & 163 & 179 & 189 & 200 \\
      & $\epsilon/a$ & \multicolumn{11}{c}{3.36}\\
      \midrule
      \multirow{2}{*}{$36^3\times 12$} & $T$ (MeV) & 115 & 122 & 125 & 132 & 139 & 140 & 145 & 149 & 152 & 158 & 160 & 167 & 178\\
      & $\epsilon/a$ & \multicolumn{11}{c}{4.32}\\
      \midrule
      \multicolumn{13}{c}{$T\sim0$ ensemble} \\
      \midrule
      \multirow{2}{*}{$24^3\times 32$} & $T$ (MeV) & 21 & 29 & \\
      & $\epsilon/a$ & \multicolumn{11}{c}{2.88}\\
      \midrule
      \multirow{2}{*}{$32^3\times 48$} & $T$ (MeV) & 27 \\
      & $\epsilon/a$ & \multicolumn{11}{c}{3.84}\\
      \midrule
      \multirow{2}{*}{$48^3\times 64$} & $T$ (MeV) & 31 \\
      & $\epsilon/a$ & \multicolumn{11}{c}{5.76}\\ 
    \bottomrule
  \end{tabular}
  \caption{\small Simulation parameters for the valence-current setup, where we kept the ratio $\epsilon/L$ fixed at 0.12. In this case, $\epsilon/a = 0.12N_s$ only depends on the number of sites in the spatial direction.}
  \label{tab:valence_setup2}
\end{table}
\section{Renormalization scheme}\label{sec:renormalization}
In this section, we will discuss the renormalization of our observables.
It is well known that the magnetic susceptibility contains an additive divergence proportional to the logarithm of the lattice spacing. This divergence stems from the multiplicative renormalization of the bare electric coupling~\cite{Schwinger:1951nm,Bali:2014kia}. Since the divergence in $\chi_m$ is temperature-independent, to obtain a finite result we can define the renormalized susceptibility as
\begin{equation}
\chi^R_m(T) \equiv \chi_m(T)-\chi_m(T=0)\,.
\label{eq:renorm_susc_scheme}
\end{equation}
As pointed out in Ref.~\cite{Bali:2020bcn}, this scheme is compatible with the requirement that the magnetic permeability $\mu$ equals unity in the vacuum. From Eq.~\eqref{eq:ampere_law}, it is clear that  the induced electric current inherits the divergence from the susceptibility. This is not what one would a priori expect from a conserved current in a quantum field theory since these, in general, avoid renormalization, with only few exceptions (see, for instance, Ref. \cite{collins2006renormalization}). Nevertheless, this is indeed the case for the electric current operator in consideration. To clarify why, let us go back to the separation~\eqref{eq:total_j} of the total current into free and magnetization currents.

Suppose that our system is subject to an external magnetic field $\vec{H}$, induced by the free current $\vec{j}_{H}$. The medium responds to the external field with a magnetization $\vec{M}$, carried by the electric current $\vec{j}$. An external observer, however, does not detect $\vec{j}_{H}$ or $\vec{j}$ individually, but only the total current $\vec{j}_{B}$, which indeed does not renormalize. By separating the medium response $\vec{j}$ from the free current $\vec{j}_{H}$, one finds ultraviolet divergent terms in both, nevertheless, they mutually cancel and give a finite total current. This means that although $\vec{j}$ is a conserved current, it is infinite in the ultraviolet due to the fact that we separated it from the free current.

Having clarified this aspect, we proceed to find the renormalization prescription for the induced current. Using Eq.~\eqref{eq:current_convol_with_wedge} and the definition of the renormalized susceptibility, we find that
\begin{align}
\int_{0}^{L/2} \ddx W(x,L)\,j^R_y(x) = \int_{0}^{L/2} \ddx W(x,L)\,j_y(x) - \chi_m(T=0)\int_{0}^{L/2} \ddx W(x,L) \dfrac{\partial eB(x)}{\partial x}\,,
\end{align}
from which we conclude that
a suitable definition of the renormalized current is
\begin{equation}
j^R_y(x,B,T) \equiv j_y(x,B,T) - \chi_m(T=0)\pdv{eB(x)}{x}\,.
\label{eq:renorm_curr}
\end{equation}
Note that this form of the renormalized current is consistent with the fact that the local susceptibility in~\eqref{eq:jy_convolution} renormalizes with a contact term, $\chi(x')-L\chi_m(T=0)\cdot \delta(x')$ or, that the divergent part of the susceptibility in Fourier-space~\eqref{eq:chip1} is momentum-independent.
We stress that Eq.~\eqref{eq:renorm_curr} is not equivalent to a subtraction of the $T=0$ current. The latter would imply a subtraction to all orders in the magnetic field, while the correct renormalization only affects the term proportional to $\partial B(x)/\partial x$.

\section{Results}\label{sec:results}

In this section, we discuss our results for the electric current density, and the magnetic susceptibility derived from it. To clearly distinguish the two approaches used, we start by showing the results from the full-current in Sec.~\ref{sec:full} and later we discuss the valence-current approach in Secs.~\ref{sec:valence_fixed_eps_physical} and~\ref{sec:valence_2}.

\subsection{Full-current results}\label{sec:full}
\begin{figure}[!h]
    \centering
    \begin{subfigure}{0.49\textwidth}
    \includegraphics[width=\linewidth]{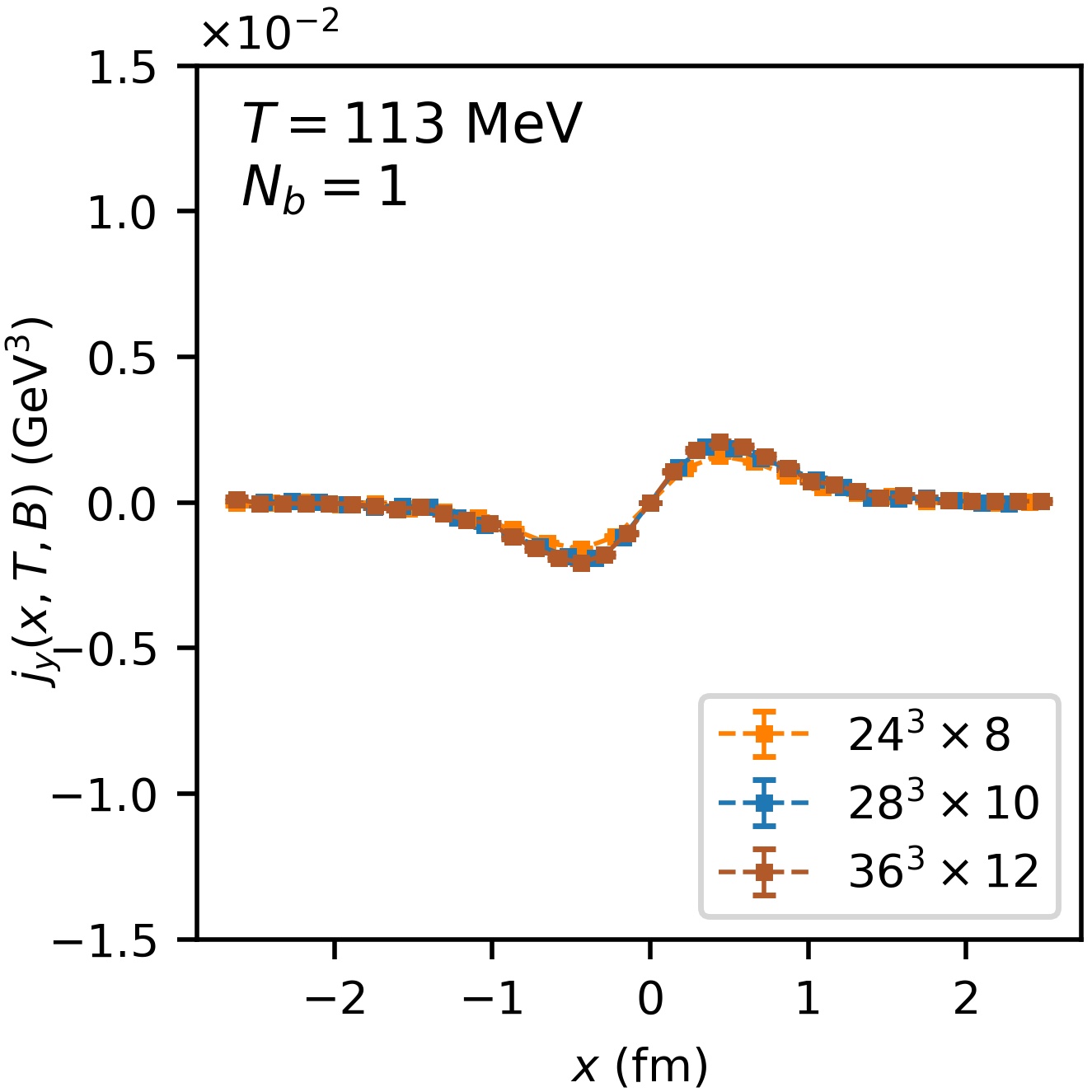}
    \end{subfigure}
    \begin{subfigure}{0.49\textwidth}
    \includegraphics[width=\linewidth]{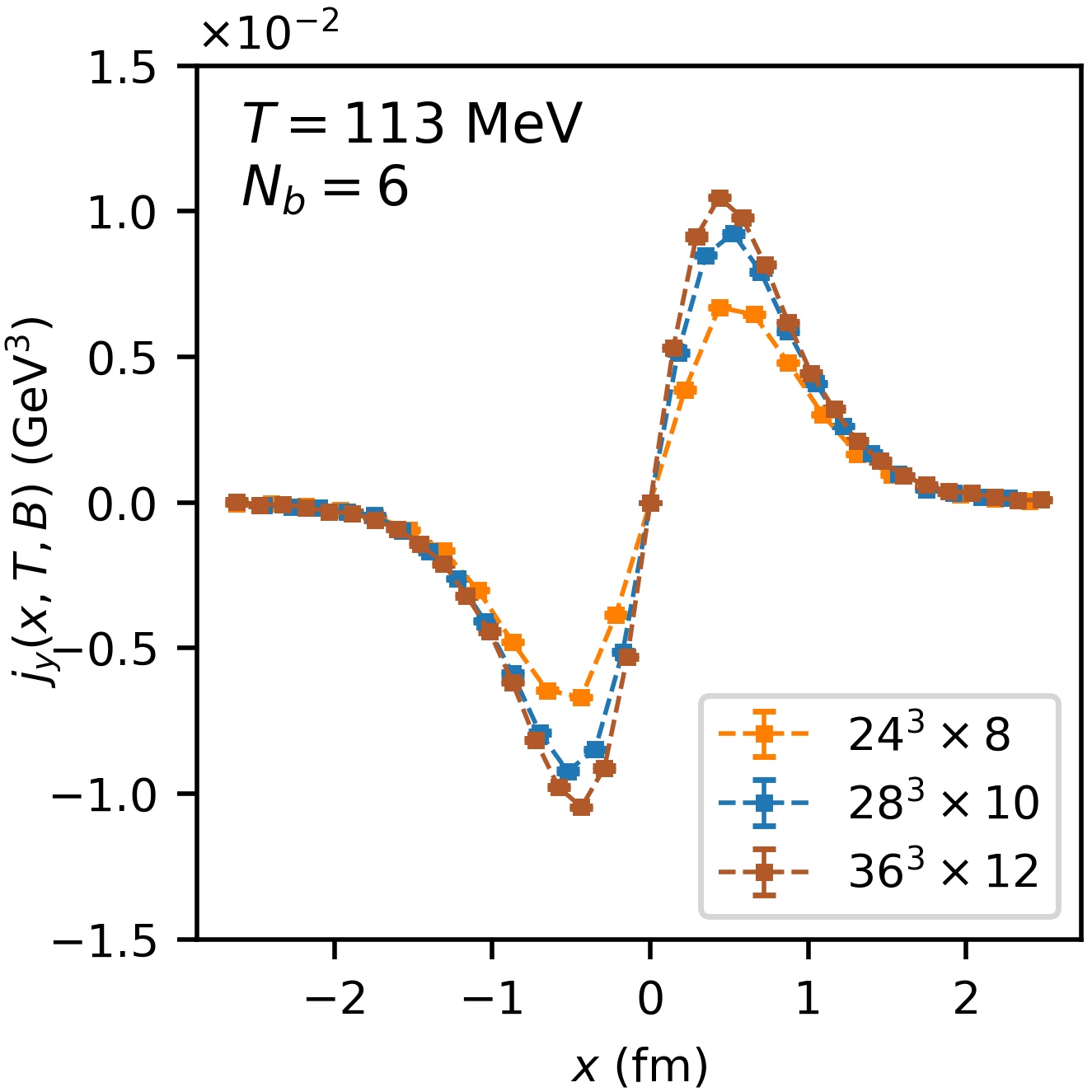}
    \end{subfigure}
    \caption{\small Left plot: bare electric current profiles as functions of $x$ at $T = 113$ MeV and $N_b=1$. Right plot: similar plot at $N_b=6$.}
    \label{fig:currents_RHIC_LHC}
\end{figure}

In Fig.~\ref{fig:currents_RHIC_LHC}, we show the bare electric current densities on the lattice computed using Eq.~\eqref{eq:staggered_current} at a given temperature for two magnetic field strengths.

Note that, from Eq.~\eqref{eq:quantization_aspec_rat}, the physical magnetic field $eB$ is different for the ensembles with slightly different aspect ratios. In particular, for the $28^3\times10$ lattice, which has aspect ratio 2.8 instead of 3, the physical magnetic field is slightly higher than the other two lattices.

\begin{figure}[!h]
    \centering
    \includegraphics{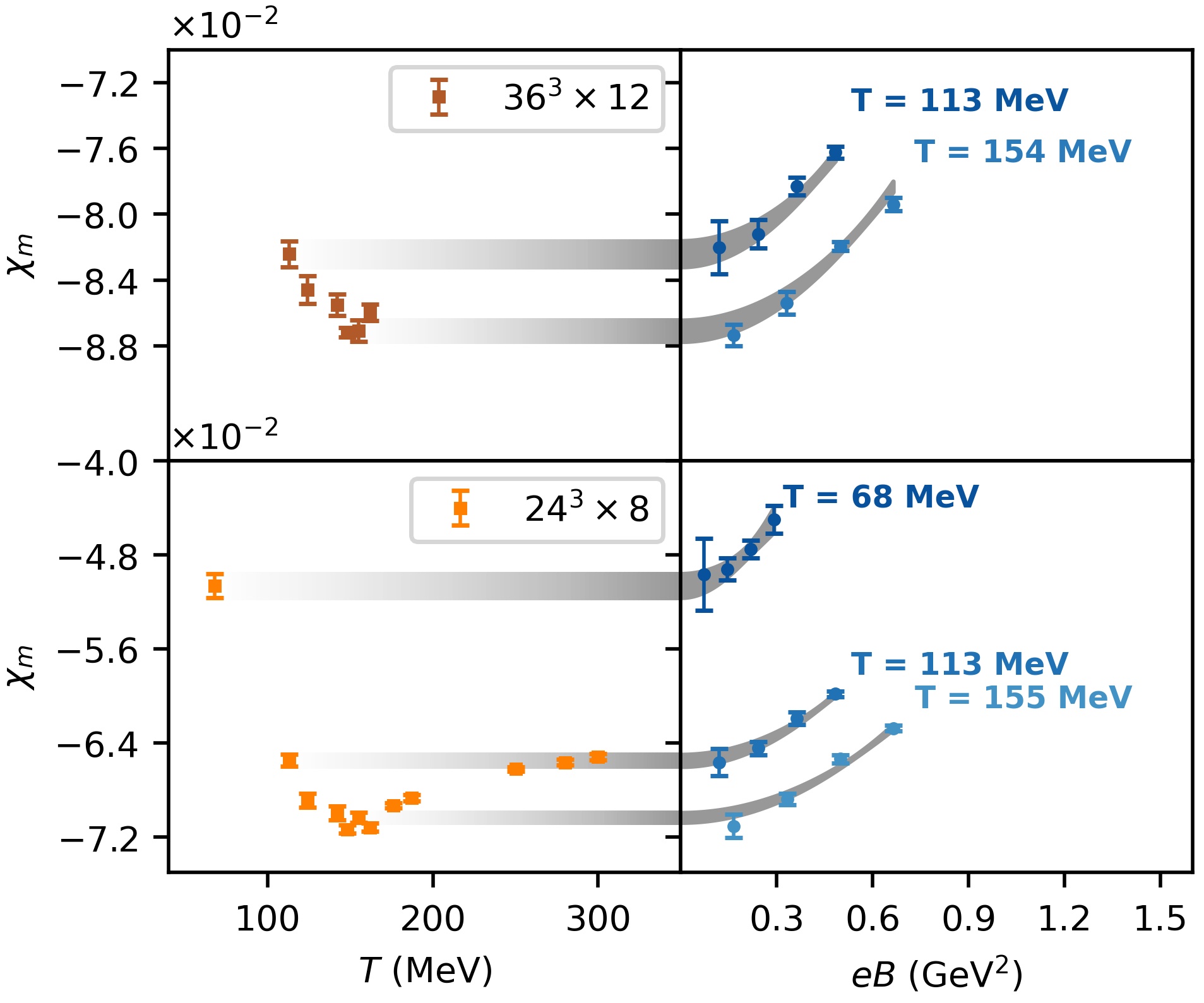}
    \caption{\small Right column plots: $\chi_m$ as a function of $eB$ obtained from Eq.~\eqref{eq:current_convol_with_wedge} with quadratic extrapolation (gray bands) for different temperatures and two lattice spacings. The width of the bands depict the statistical error of the fits to the data on the right-hand-side. Left column plots: projection of the zero-magnetic-field $\chi_m$ on the temperature axis for two lattice sizes. The gray bands, representing the statistical error of the fits, were extended to the $T$-axis to guide the eye.}
    \label{fig:all_fits}
\end{figure}

As discussed in Sec.~\ref{sec:mag_susc}, these profiles can be combined with Amp\`{e}re's law to obtain the magnetic susceptibility. The magnetic flux quantization only allows us to access discrete values of $N_b$. Therefore, to carry out the zero-$B$ limit, we fitted a quadratic function to the ratio in Eq.~\eqref{eq:current_convol_with_wedge}. In Fig.~\ref{fig:all_fits}, we illustrate the extrapolation of the bare magnetic susceptibility to $B=0$ for two lattice spacings, as described in Sec.~\ref{sec:mag_susc}, and the corresponding temperature dependence. 

For some temperatures, we found that the data points at $N_b = 1$ show a slight deviation from the expected quadratic behavior. This issue can be explained in the following way: it has been shown that finite-volume effects induce sinusoidal modulations in some thermodynamic quantities even at uniform background magnetic field \cite{adhikari2023qcd}, where the frequency of the oscillation increases with $N_b$. In particular, we found this to also hold for the electric current. At $N_b = 1$, the finite-volume contamination overlaps with the wedge function in Eq.~\eqref{eq:current_convol_with_wedge}, and its signal is therefore enhanced by it. This, however, was found to be unproblematic for the $B\to0$ extrapolations and did not have a substantial impact on the renormalized susceptibility.

\begin{figure}[!h]
    \centering
    \includegraphics{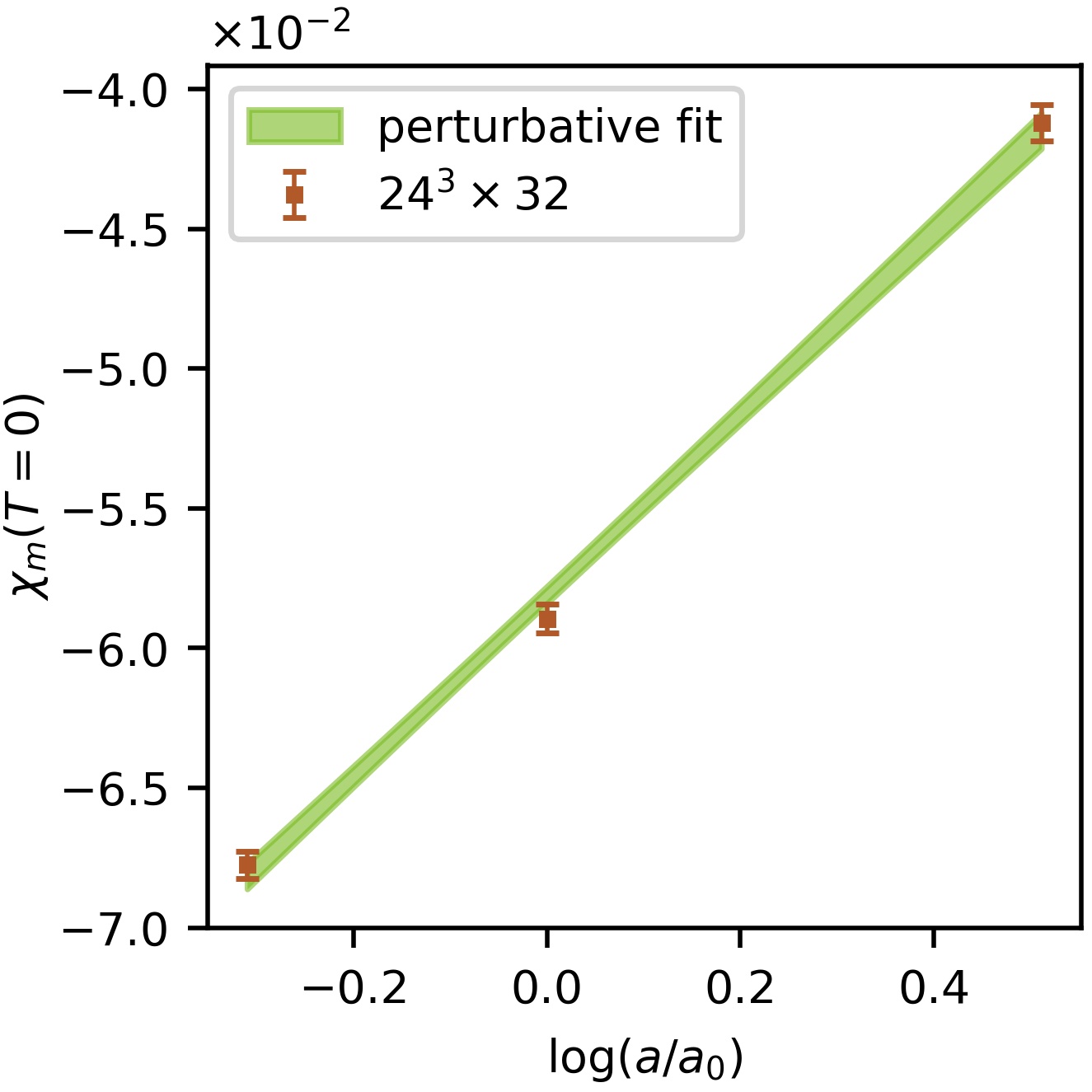}
    \caption{\small Bare magnetic susceptibility at zero-temperature and zero-field obtained in the full-current setup with linear perturbative fit as a function of $\log(a/a_0)$. $a$ is the lattice spacing and $a_0 \equiv 1.46$ GeV$^{-1}$ is a reference lattice spacing, as defined in Ref.~\cite{Bali:2020bcn}.}
    \label{fig:susc_T0}
\end{figure}

To renormalize the current and the susceptibility, we needed to carry out the extrapolation to $B=0$ for the $T=0$ ensemble and subtract the corresponding divergent contribution. In Fig.~\ref{fig:susc_T0}, we show the result of Eq.~\eqref{eq:current_convol_with_wedge_B0_limit} at $T=0$ and the linear fit in $\log(a/a_0)$, where $a$ is the lattice spacing and $a_0 \equiv 1.46$ GeV$^{-1}$ is a normalization factor. 

\begin{figure}[!h]
    \centering
    \includegraphics{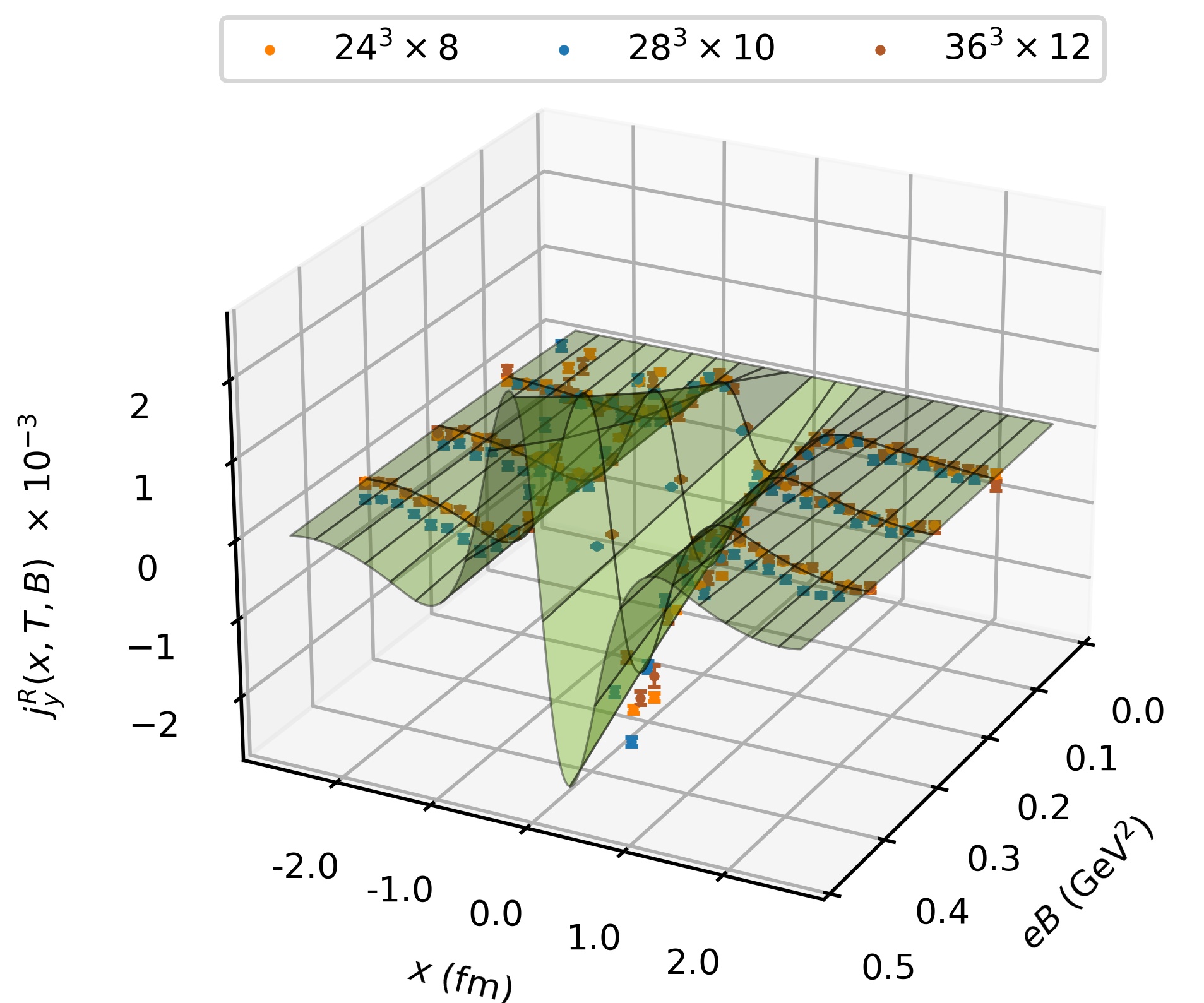}
    \caption{Continuum limit surface and renormalized electric current data at $T = 113$ MeV for $N_b = 1,2,3$. The higher $N_b$ values were included in the global fit but omitted from the plot for clarity.}
    \label{fig:cont_lim_2d}
\end{figure}

\begin{figure}[!h]
    \centering
    \begin{subfigure}{0.49\textwidth}
    \includegraphics[width=\linewidth]{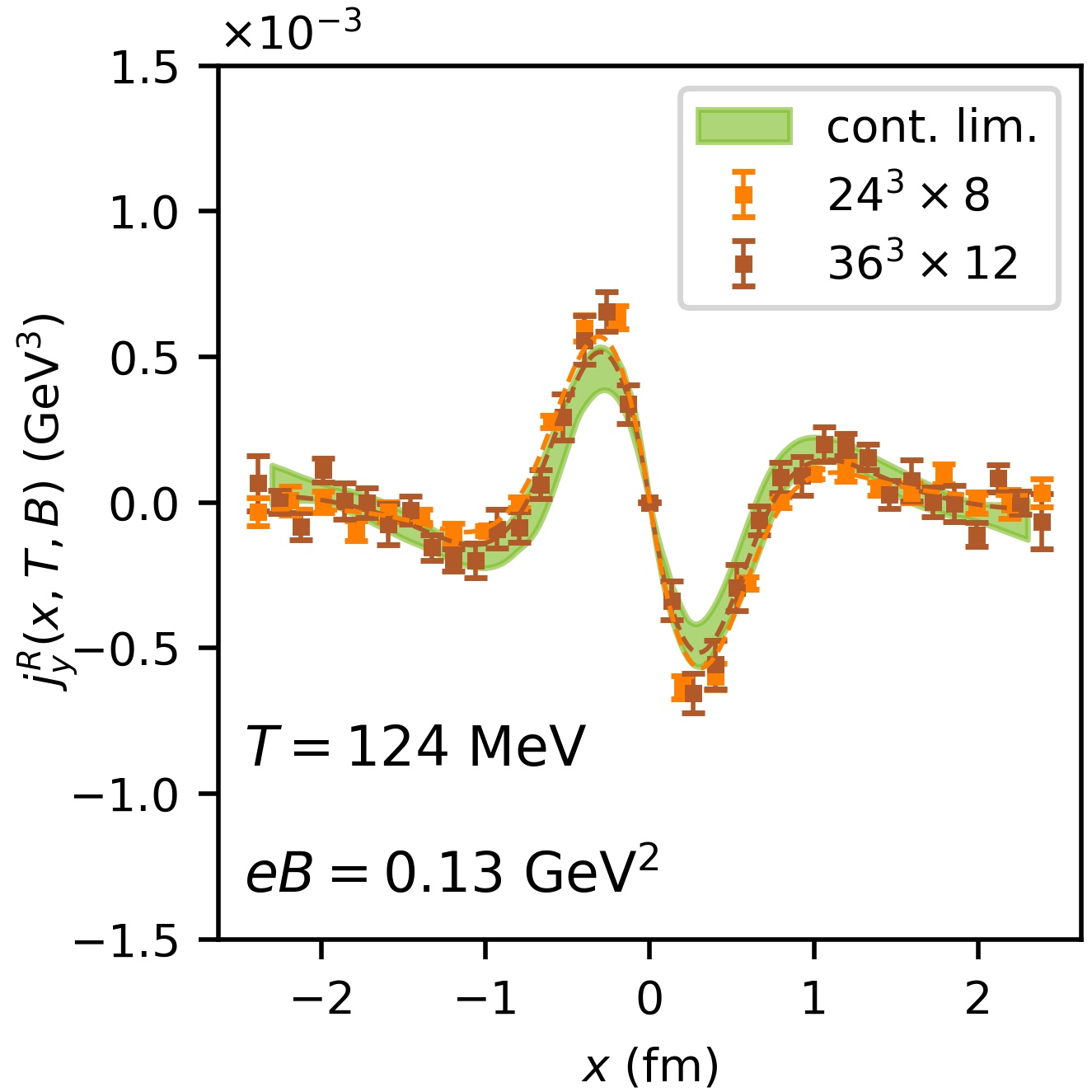}
    \end{subfigure}
    \begin{subfigure}{0.49\textwidth}
    \includegraphics[width=\linewidth]{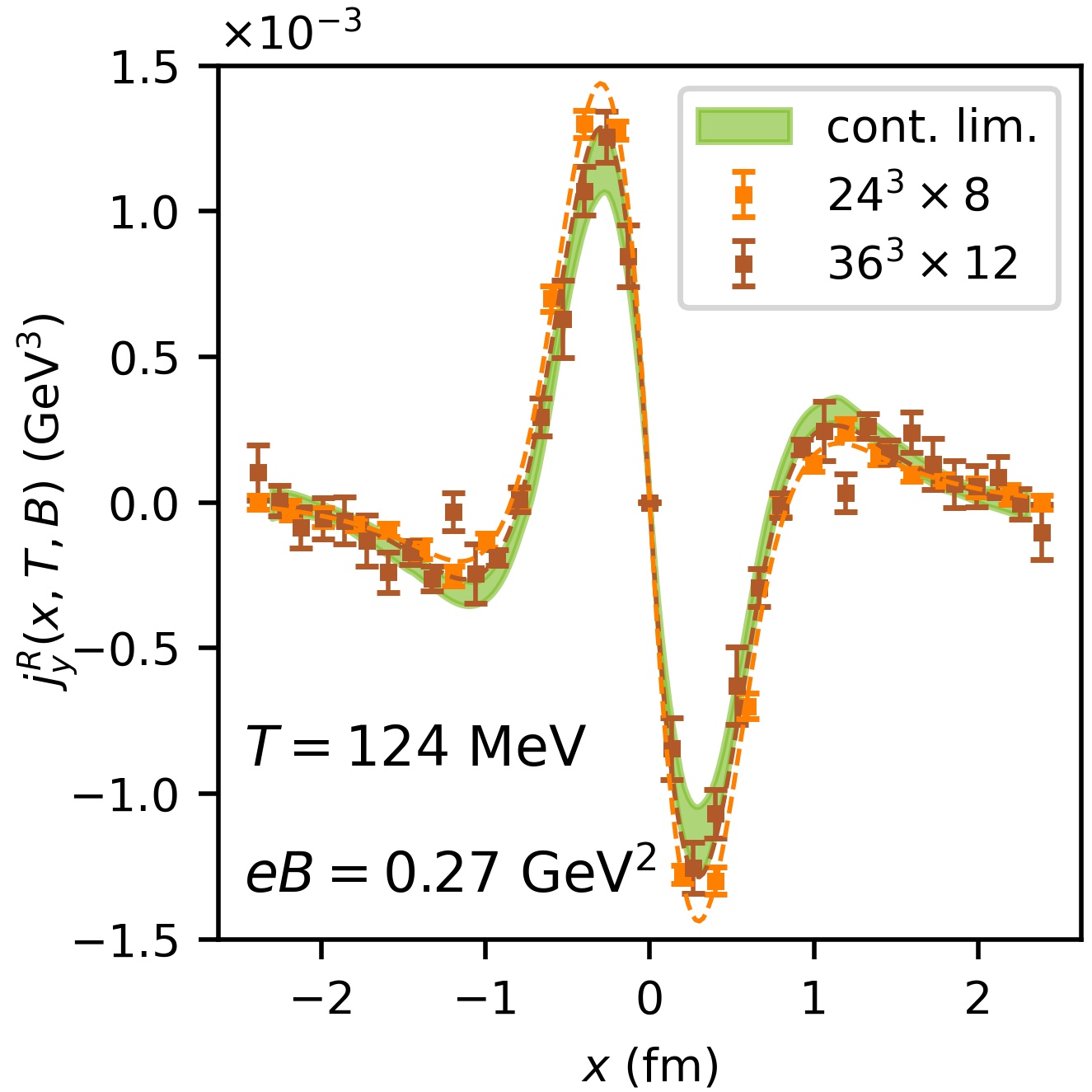}
    \end{subfigure}
    \caption{\small Continuum limit of the renormalized electric current profiles projected on the $x$ direction for two magnetic fields at $T=124$ MeV. The points represent two lattices used in the multidimensional continuum extrapolation. The dashed lines are the results of the splines at the corresponding lattice spacing.}
    \label{fig:renorm_curr}
\end{figure}

Using $\chi_m(T=0)$ obtained from the linear fit, we renormalized the current as described in Sec.~\ref{sec:renormalization}. Our continuum limit extrapolation was done by fitting the lattice data in $x$, $B$, and the lattice spacing simultaneously using a Monte Carlo approach. We carried out this procedure for the chiral condensate and Polyakov loop in Ref.~\cite{Brandt:2023dir}. For more details on the method, see \cite{Endrodi:2010ai}. In Fig.~\ref{fig:cont_lim_2d}, we show the continuum limit surface fitted to the data to illustrate the continuum extrapolation procedure. In Fig.~\ref{fig:renorm_curr}, we show slices of the continuum limit surface as a function of $x$ at specific magnetic field strengths. In Appendix~\ref{app:chiPT}, we also compute the induced electric current in chiral perturbation theory from the momentum-dependent susceptibility $\widetilde{\chi}(p)$ and compare it with our renormalized lattice data.

\begin{figure}[!h]
    \centering
    \includegraphics{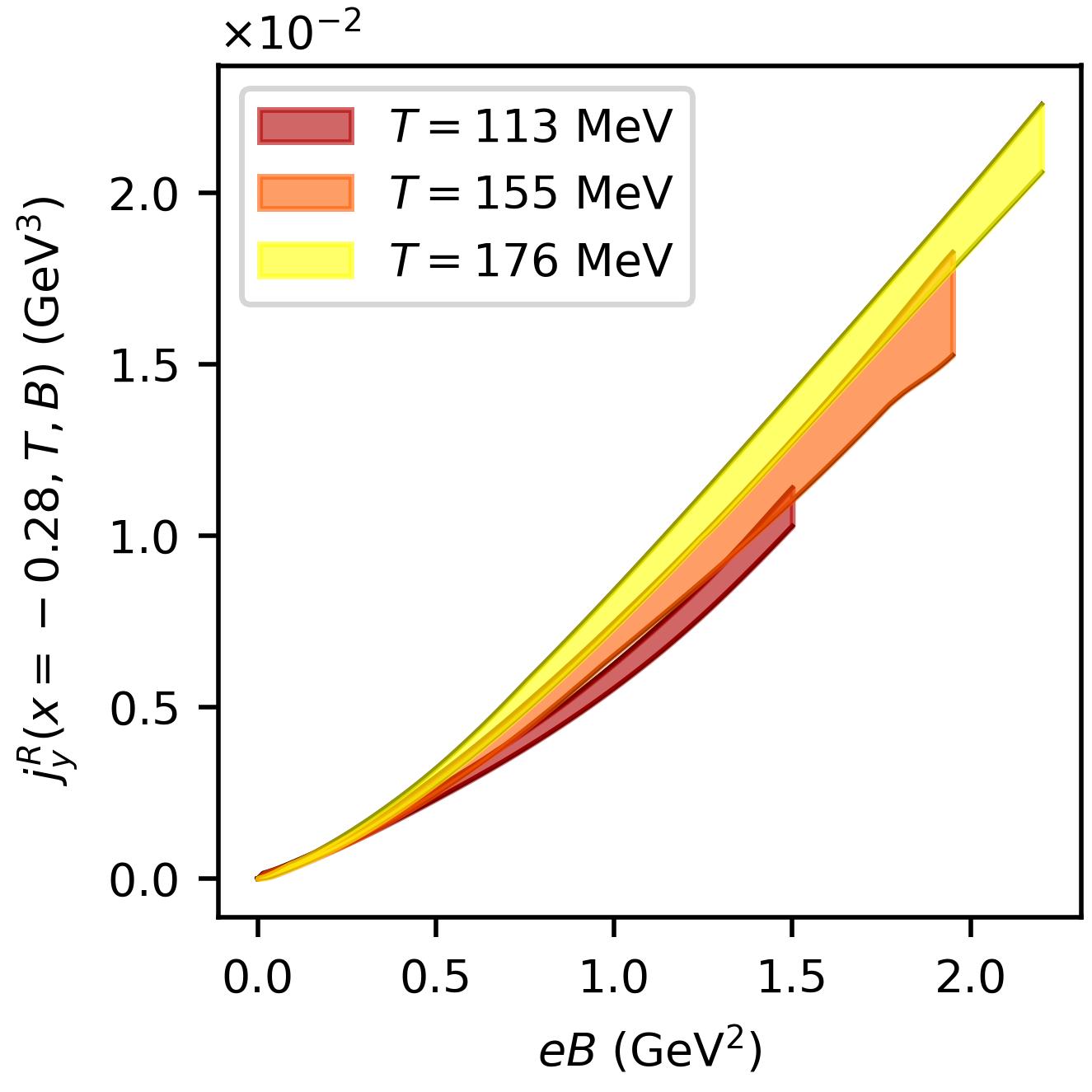}
    \caption{\small Renormalized electric currents evaluated at $x=0.28$ fm (at the peak position) in the continuum limit as a function of the magnetic field for three temperatures.}
    \label{fig:current_peak_B_field}
\end{figure}

Notice that the peak of the electric current appears around $x\sim\pm0.28$ fm, where the curl of the magnetic field is maximal. In Fig.~\ref{fig:current_peak_B_field}, we show the behavior of this peak as a function of the magnetic field for various temperatures.

\begin{figure}[!h]
    \centering
    \includegraphics{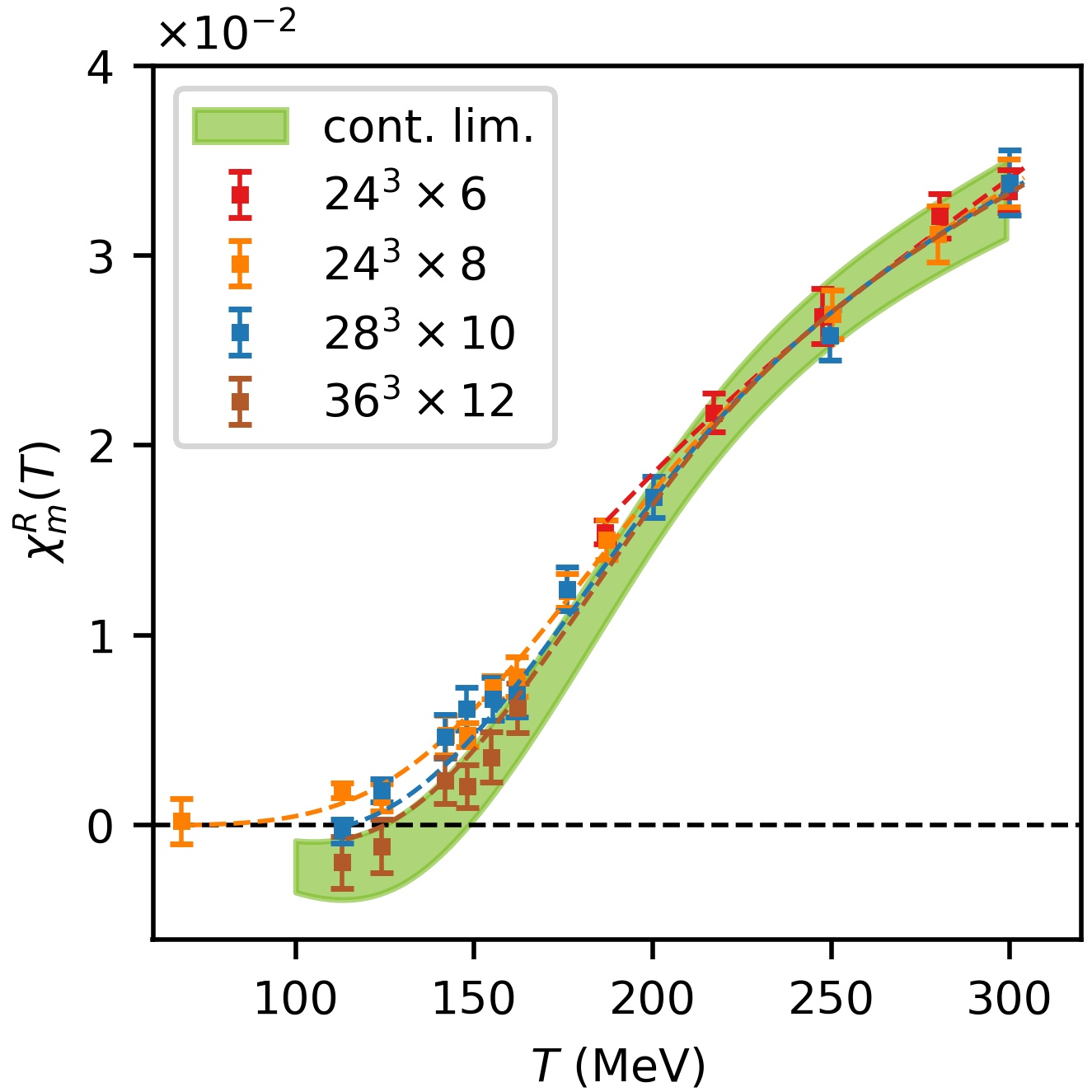}
    \caption{\small Continuum limit extrapolation of the magnetic susceptibility (green band) in the full-current setup as a function of temperature. The colored points represent the data for different lattice spacings.}
    \label{fig:mag_susc_full}
\end{figure}

In Fig.~\ref{fig:mag_susc_full}, we show the renormalized susceptibilities at nonzero lattice spacings and their continuum limit extrapolation.
At high $T$, we see that $\chi_m^R > 0$, indicating a strong paramagnetic phase, whereas at $T < T_c$, the sign of $\chi_m^R$ changes and we observe a weak diamagnetic behavior. This is in agreement with the lattice results obtained with a different method~\cite{Bali:2020bcn}.

\subsection[Valence-current results at fixed $\epsilon$]{Valence-current results at fixed \boldmath $\epsilon$}\label{sec:valence_fixed_eps_physical}

\begin{figure}[!h]
    \centering
    \begin{subfigure}{0.49\textwidth}
    \includegraphics[width=\linewidth]{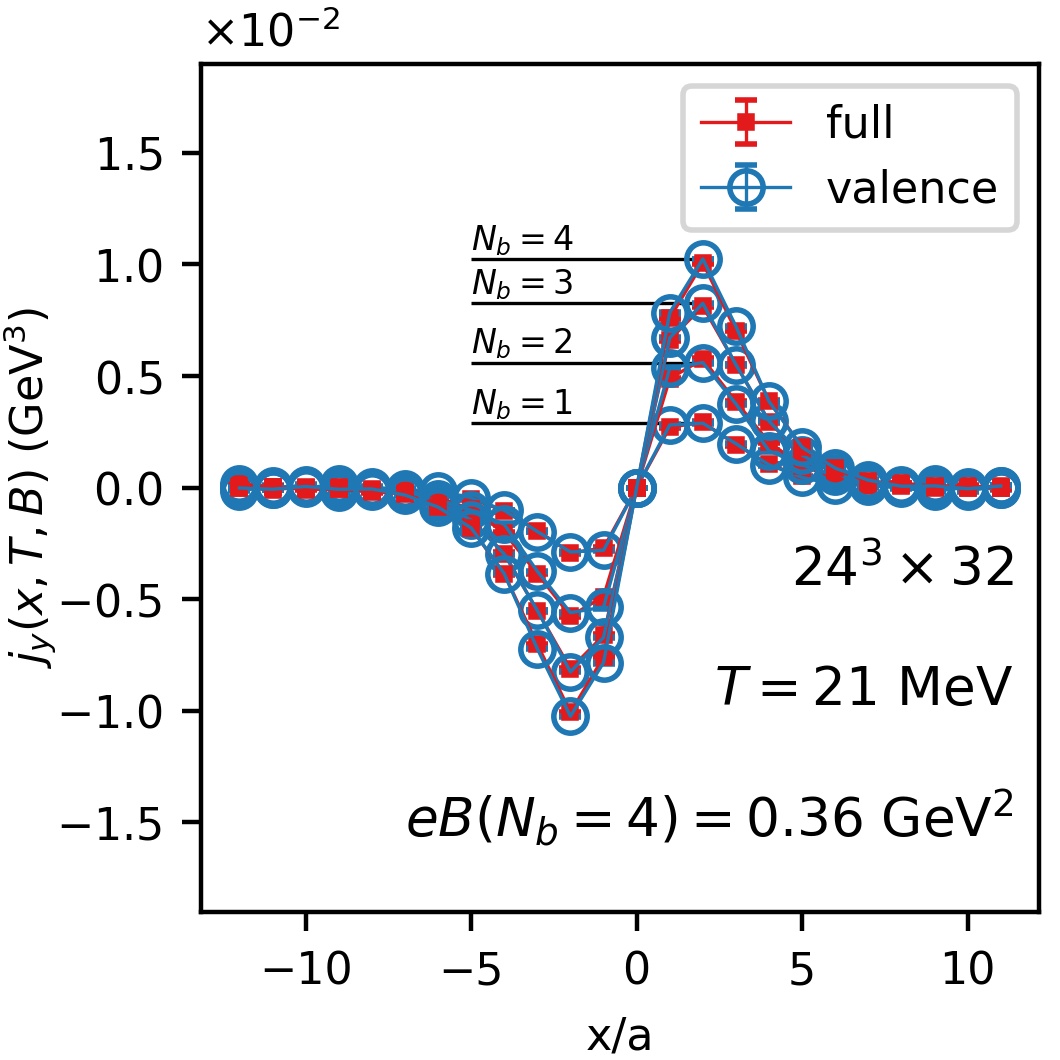}
    \end{subfigure}
    \begin{subfigure}{0.49\textwidth}
    \includegraphics[width=\linewidth]{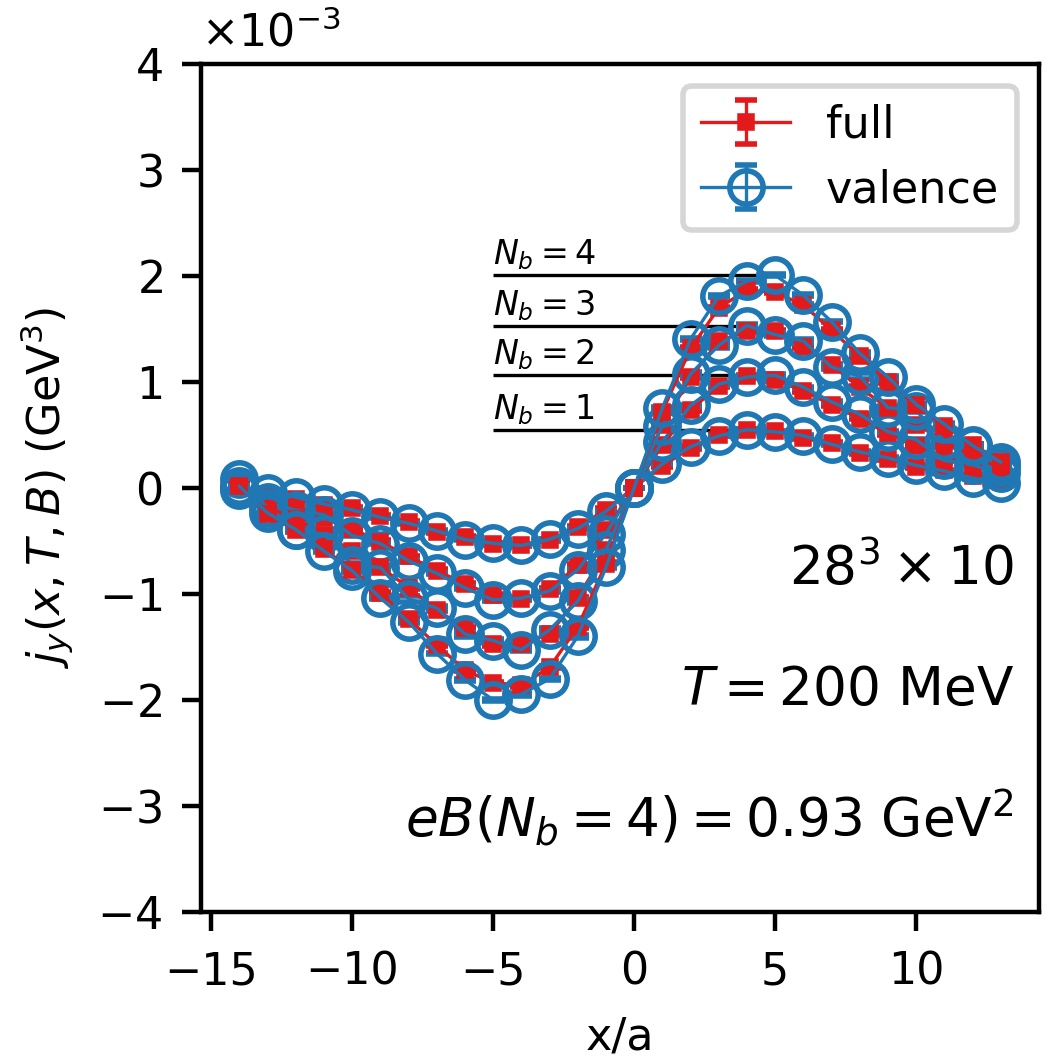}
    \end{subfigure}
    \caption{\small Comparison between valence and full currents for various magnetic fields at temperatures $T < T_c$ and $T > T_c$. In each plot, we indicate the magnetic field strength, in physical units, corresponding to the highest $N_b$ shown in the plot, i.e.\ $N_b=4$.}
    \label{fig:val_vs_full}
\end{figure}

In this section, we present our results employing the sea-valence decomposition, as defined in Sec.~\ref{sec:valence}. In Fig.~\ref{fig:val_vs_full}, we compare the magnitude of the unrenormalized valence current with the full one for different temperatures and magnetic fields.

\begin{figure}[!h]
    \centering
    \begin{subfigure}{0.49\textwidth}
    \includegraphics[width=\linewidth]{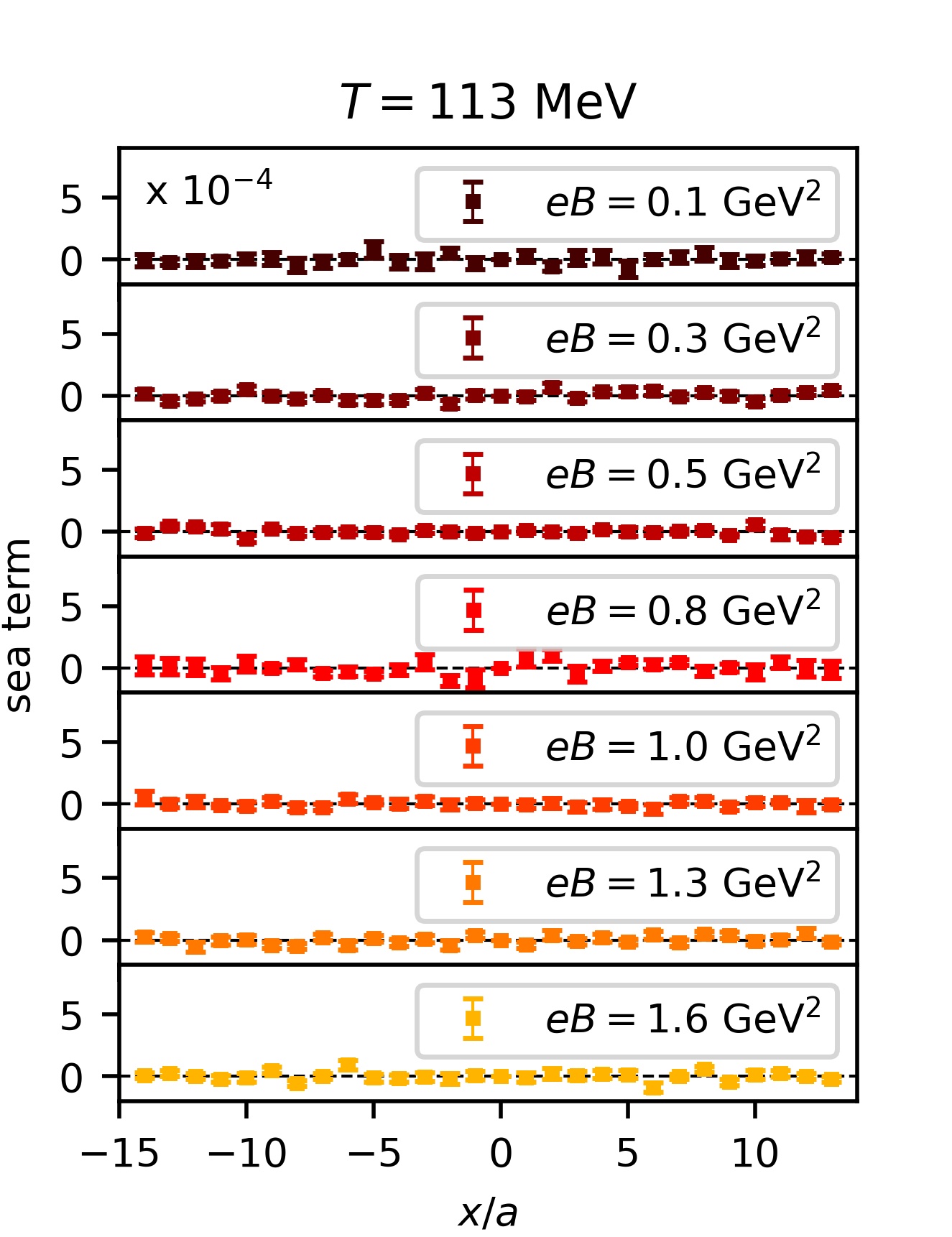}
    \end{subfigure}
    \begin{subfigure}{0.49\textwidth}
    \includegraphics[width=\linewidth]{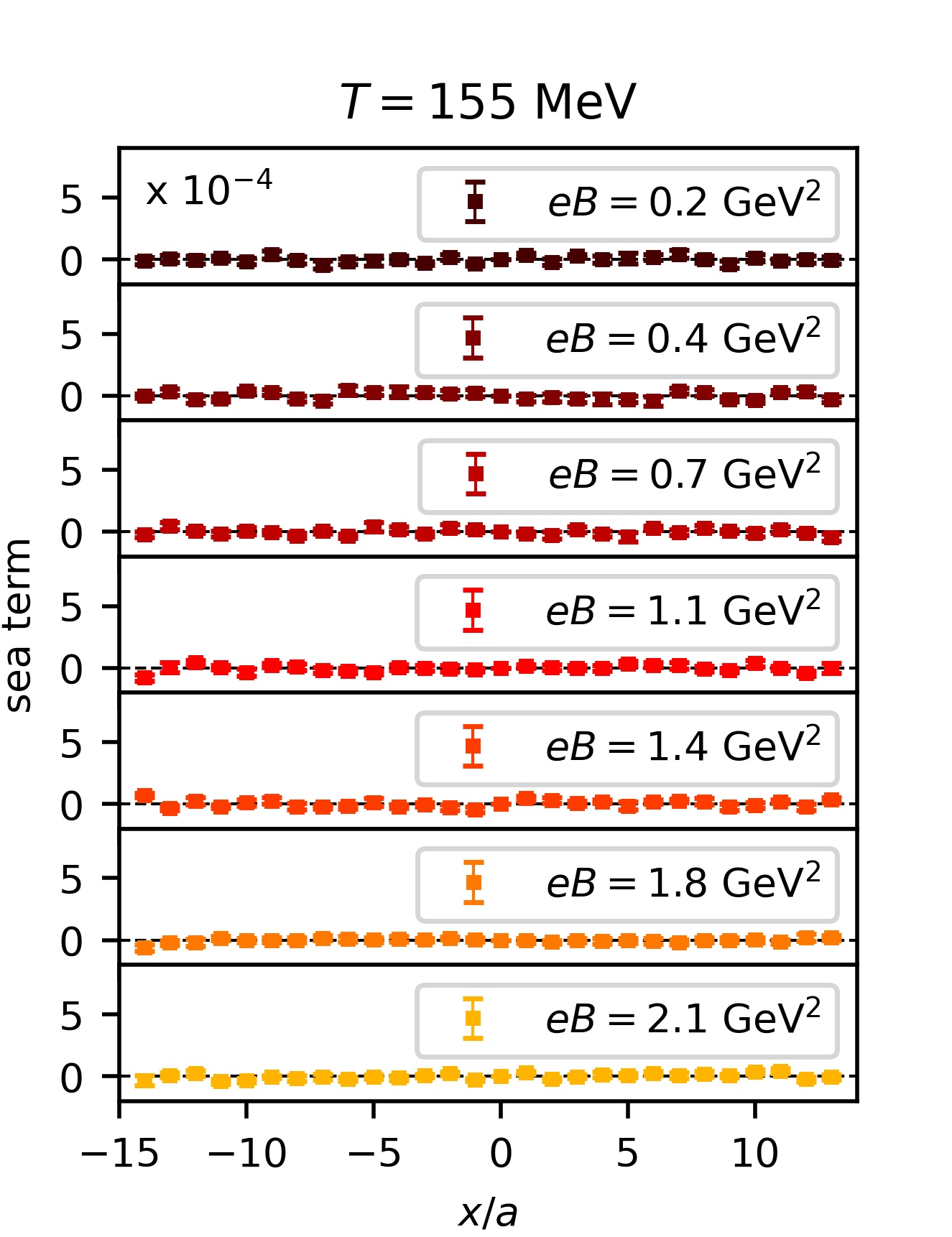}
    \end{subfigure}
    \caption{\small Sea contributions to the electric current density at $T = 113$ MeV (left) and at $T = 155$ MeV (right) as functions of the $x$-coordinate for different magnetic field strengths on a $28^3\times10$ lattice. The sea term fluctuates closely to zero even in the strong-field regime.}
    \label{fig:sea_curr}
\end{figure}

In Fig.~\ref{fig:sea_curr}, we also show the bare sea contribution to the current for various magnetic field values on one particular lattice.

\begin{figure}
    \centering
    \begin{subfigure}{0.49\textwidth}
    \includegraphics[width=\linewidth]{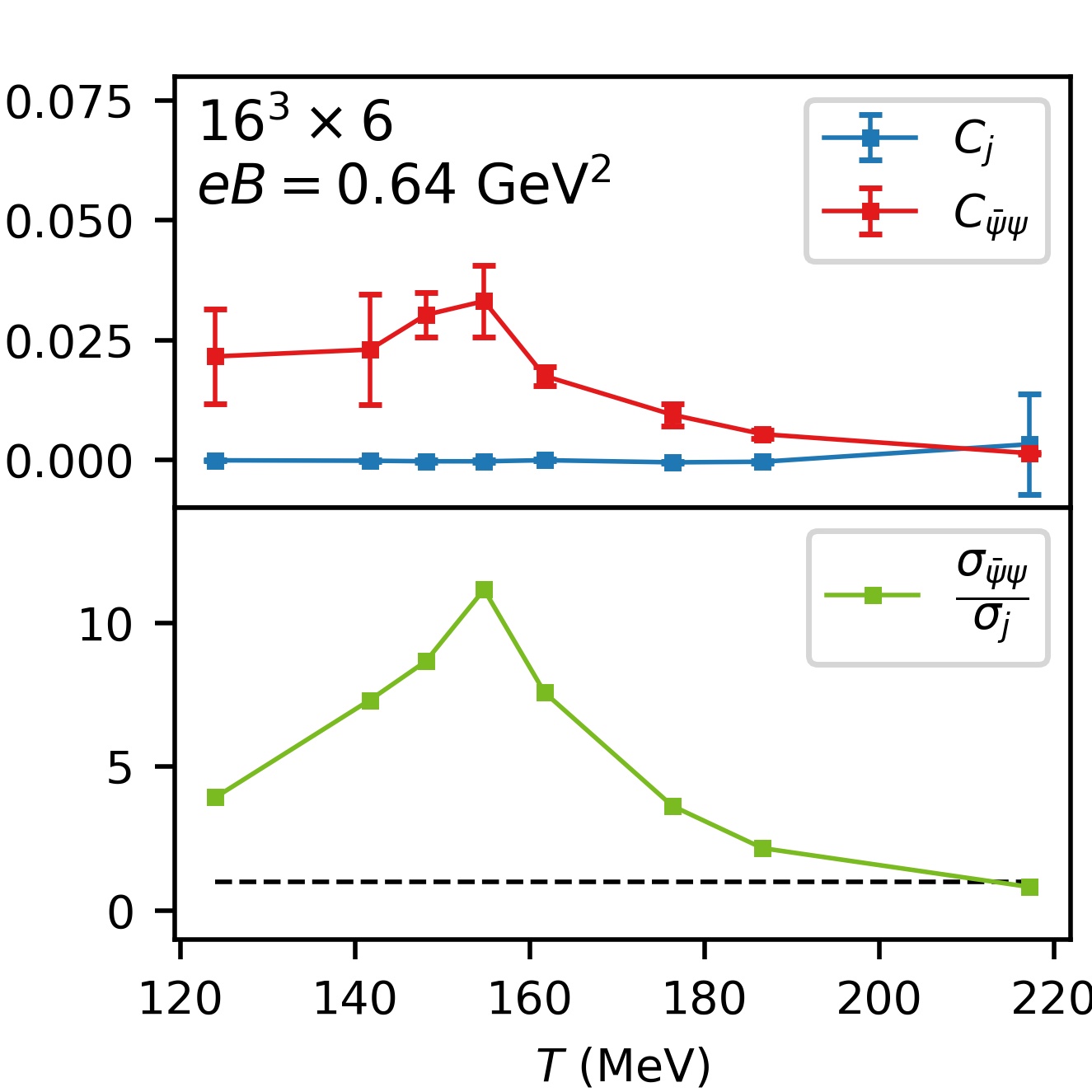}
    \end{subfigure}
    \begin{subfigure}{0.49\textwidth}
    \includegraphics[width=\linewidth]{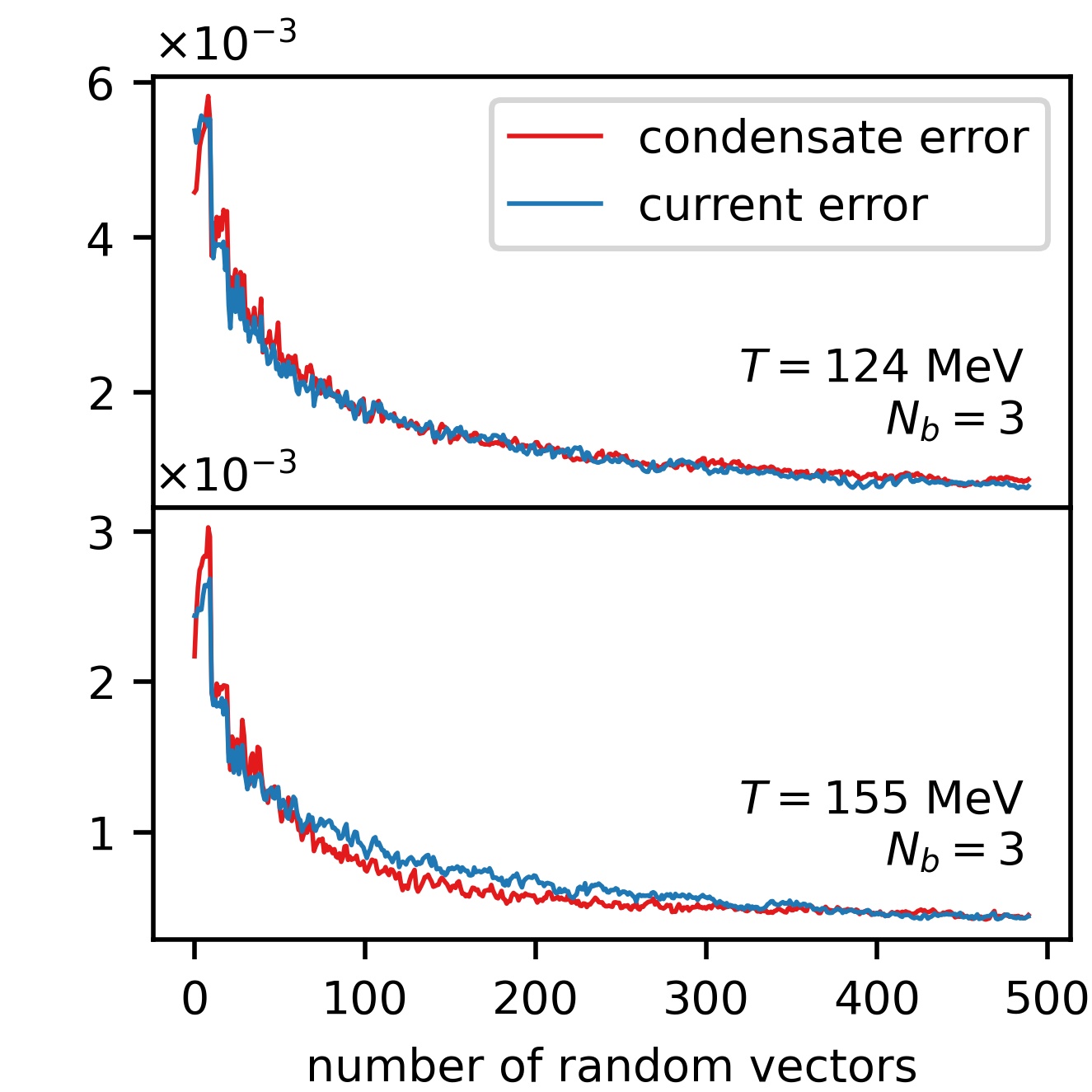}
    \end{subfigure}
    \caption{\small Upper left panel: correlators between the Polyakov loop and the electric current (blue) and between the Polyakov loop and the chiral condensate (red) as functions of $T$. Lower left panel: ratio between the average statistical errors for the chiral condensate ($\sigma_{\bar{\psi}\psi}$) and the electric current ($\sigma_j$) on this ensemble. The dashed line represents $\sigma_{\bar{\psi}\psi}/\sigma_{j} = 1$. Right panels: average error coming from the random vectors on one gauge configuration for the current and chiral condensate at $T = 124$ MeV and $T = 155$ MeV as a function of the number of random vectors.}
    \label{fig:ploop_current_corr}
\end{figure}

Figs.~\ref{fig:val_vs_full} and~\ref{fig:sea_curr} provide empirical evidence that the valence term captures most of the physics of the electric current in the weak-field regime. Since the sea term quantifies the effect of quark loops in gluonic interactions at low $B$, its tiny magnitude suggests that the electric current operator is very weakly sensitive to the gluon fluctuations. To further quantify this effect, we define a correlator between the bare electric current and the bare Polyakov loop. The latter operator encodes the most relevant properties of the gluon fields in the thermal medium. The correlator reads

\begin{equation}
C_{j} = -\frac{1}{m_{\pi}^3}\sum_{x=0}^{L/2}\sum_{x^{\prime}=0}^{L/2} \qty[\ave{j_y(x)P(x^{\prime})} - \ave{j_y(x)}\ave{P(x^{\prime})}]\,,
\end{equation}
where we normalized by the cubed pion mass to have a dimensionless combination. The integration is carried out on half of the $x$-direction so that $C_j$ is non-zero. This correlator is shown in Fig.~\ref{fig:ploop_current_corr} to demonstrate that it vanishes within errors. 

For comparison, we can define the analogous correlator between the Polyakov loop and the chiral condensate (see~\cite{Brandt:2023dir}).
\begin{equation}
C_{\bar{\psi}\psi} = -\frac{1}{m_{\pi}^3}\sum_{x=0}^{L/2}\sum_{x^{\prime}=0}^{L/2} \qty[\ave{\bar{\psi}\psi(x)P(x^{\prime})} - \ave{\bar{\psi}\psi(x)}\ave{P(x^{\prime})}]\,,
\end{equation}
where $\bar\psi\psi$ is the average over the light-quark condensates. Fig.~\ref{fig:ploop_current_corr} shows that the magnitude of $C_{\bar\psi\psi}$ is substantially larger, peaking around the transition temperature where correlations are most prominent.
We note moreover that the low sensitivity of the current to gluons also manifests itself in a reduction of statistical uncertainties associated with this operator on a given ensemble. For instance, we found that the magnitude of the errors $\sigma$ are typically much lower in the case of the current as compared to the chiral condensate, which has a non-vanishing sea contribution. In Fig.~\ref{fig:ploop_current_corr}, we also include a comparison between $C_j$ and $C_{\bar{\psi}\psi}$ and the ratio $\sigma_{\bar{\psi}\psi}/\sigma_j$ as functions of $T$.
For concreteness, we note that for the same ensemble, the condensate can have up to $\sim10$ times larger statistical errors than the current. When analyzed on a single gauge configuration, $\sigma_j$ and $\sigma_{\bar{\psi}\psi}$ behave rather similarly as a function of the number of random vectors, indicating that, indeed, the factor of 10 difference between them is better explained by the gauge noise.

\subsection{Valence-current results at fixed $\epsilon/L$}\label{sec:valence_2}

\begin{figure}[!h]
    \centering
    \begin{subfigure}{0.49\textwidth}
    \includegraphics[width=\linewidth]{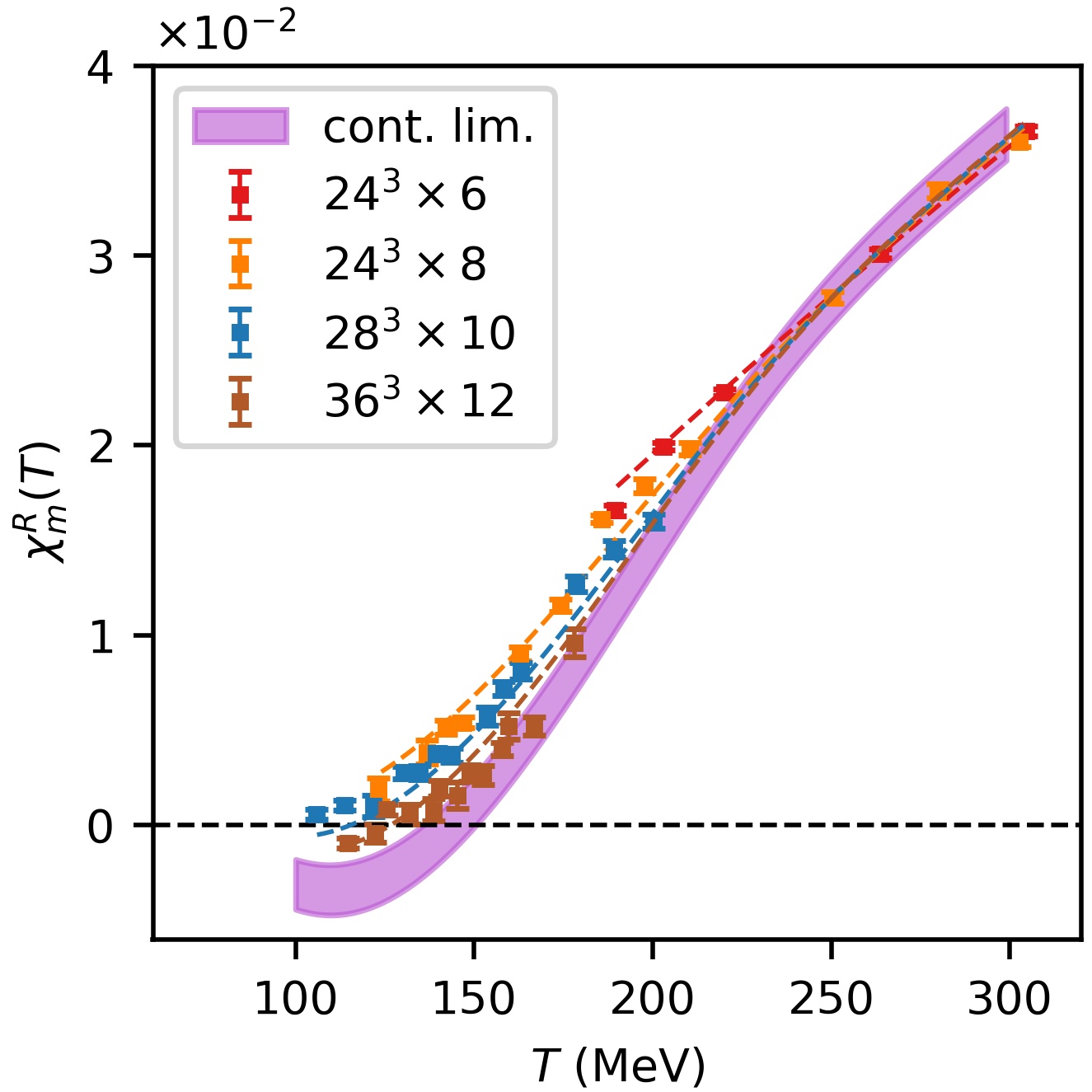}
    \end{subfigure}
    \begin{subfigure}{0.49\textwidth}
    \includegraphics[width=\linewidth]{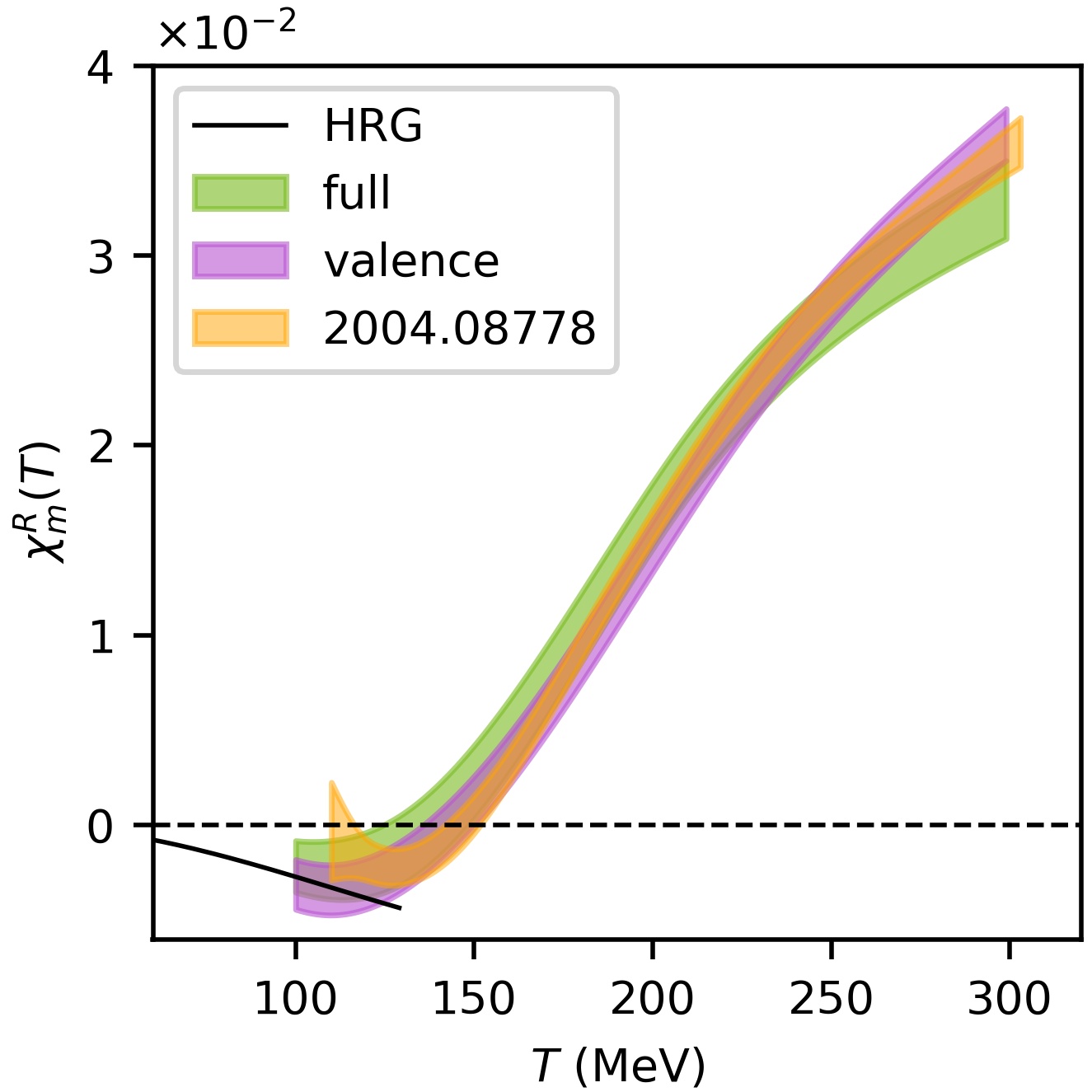}
    \end{subfigure}
    \caption{\small Left plot: continuum limit extrapolation of the magnetic susceptibility (purple band) in the valence setup as a function of temperature. The points represent the lattice data. Right plot: comparison between the magnetic susceptibility estimated using the full-current and the valence-current methods. In addition, the continuum extrapolated results of~\cite{Bali:2020bcn} using an alternative method (yellow band), as well as the result of the HRG model~\cite{Bali:2014kia} (black line) are also included.} 
    \label{fig:mag_susc_m1}
\end{figure}

The procedure for the valence-current method is entirely analogous to the one of the full-current, except that now we take $j_y$ as the valence term in Eq.~\eqref{eq:val_term}. In Fig.~\ref{fig:mag_susc_m1}, we compare the susceptibilities estimated from the full and valence currents, as well as an alternative method in the literature employing current-current correlators~\cite{Bali:2020bcn}. Here we also include the prediction of the hadron resonance gas (HRG) model~\cite{Endrodi:2013cs,Bali:2014kia}.

\subsection{Discussion}
In the previous section, we investigated the appearance of an electric current profile at non-uniform magnetic fields using three approaches: 1. Employing the full electric current operator at fixed $\epsilon$, 2. Making use of the valence approximation at fixed $\epsilon$, and 3. Making use of the valence approximation at fixed $\epsilon/L_x$.

The precise agreement between approaches 1 and 2 on the electric current throughout a broad range of temperatures confirms the validity of this new approximation scheme developed in this work. It also paved the way for our third approach, which was used to compute the magnetic susceptibility in the valence approximation scheme. Interestingly, the valence approximation suggests that in a magnetized medium, valence quarks are more efficient carriers of the electric current than sea quarks. As we have shown, this physical picture is supported by two of our results from the previous section: 1. the tiny magnitude of the sea contribution to the total current, as seen in Fig.~\ref{fig:sea_curr}, even at strong magnetic fields and 2. the fact that the correlator between the current and the Polyakov loop vanishes within errors. This finding also reveals the low sensitivity of the electric current to the gluonic fluctuations in the medium, which explains the smaller statistical errors on the computation of $\ave{\bar{\psi}\Gamma_{\mu}\psi}$ compared to the chiral condensate on a given ensemble. 

Another interesting behavior of the chiral condensate in the presence of magnetic fields that is completely absent in the electric current is the so-called inverse magnetic catalysis~\cite{Bruckmann:2013oba}. Around $T_c$, sea effects dominate over valence effects and the chiral condensate is known to be supressed by the field. However, as shown in Fig.~\ref{fig:current_peak_B_field}, the peak of the electric current is a monotonic function of $B$ for all temperatures, where the dependence on the temperature was also found to be mild. This is directly related to the fact that sea effects on the electric current are negligible at low $B$ and correlations with the Polyakov loop are practically absent.

At stronger $B$, however, the valence approximation breaks down. For instance, in Figure~\ref{fig:val_vs_full}, at $T = 200$ MeV, we start to see a small deviation from the full current at $eB \gtrsim 0.9$ GeV$^2$. Nevertheless, the sea term alone cannot explain the difference between the full current and the valence current, as one can see in Fig.~\ref{fig:sea_curr}. Therefore, the origin of the slight discrepancy between full current and valence current is attributed to higher-order contributions in the sea-valence expansion (see Eq.~\eqref{eq:sea_valence}).  

A question that arises is whether one can extend this approximation to other $B$-odd operators, for instance those related to anomalous transport phenomena~\cite{Brandt:2023wgf}. Moreover, due to the low computational cost of the valence current computation, the approach we developed in this work might be easily implemented within other fermion discretizations to study the weak-field regime of operators. In addition, the method may also be extended to arbitrary profiles of the magnetic background.
\section{Conclusions}
\label{sec:conclusions}
In this article, we studied the emergence of local electric currents in QCD in the presence of magnetic fields with non-zero curl using lattice simulations. We extrapolated our results to the continuum limit and determined the spatial structure of these currents, as well as their magnetic field and temperature dependence. As discussed in Sec.~\ref{sec:intro}, such currents are allowed by Bloch's theorem, as long as the their integral over the full volume vanishes, which is the case at hand. Furthermore, using the sea-valence expansion of the electric current, we empirically established that sea quark effects are negligible for its weak-field behavior. This lead us to the valence approximation as a computationally comparably cheap technique. Furthermore, we determined that the sea contribution to the electric current remains very close to zero even at strong fields. Therefore, we attributed the slight discrepancy between the full current and the valence current to higher-order terms in the sea-valence expansion. 

Employing the full current operator, and also the valence approximation, we showed two novel ways of computing the magnetic susceptibility of QCD applying Amp\`{e}re's law. 
This method draws a direct connection between the susceptibility (a two-point function) and the electric current (a one-point function), namely Eq.~\eqref{eq:current_convol_with_wedge}, which is only possible due to the inhomogeneity of the field. In this sense, the current-current correlator method of~\cite{Bali:2015msa,Bali:2020bcn} may be viewed as the exact weak-field expansion of the local currents in our approach.

The magnetic susceptibility obtained from the full-current method and from the valence approximation agree with the results obtained in~\cite{Bali:2020bcn} on the quantitative level. We thereby confirmed that the susceptibility changes sign across the QCD crossover, going from diamagnetic to paramagnetic matter, which pinpoints the transition from a system dominated by hadrons to a system with quasi-free quarks. 

In the context of off-central heavy-ion collision phenomenology, the benchmarking of the initial magnetic 
field strength has recently shifted to the focus
of interest, with various suggestions for observables 
that may be measured both experimentally as well as 
in lattice simulations, see e.g.\ Ref.~\cite{Ding:2023bft}.
The induced currents that we explored in this paper are 
paramagnetic at high temperatures, i.e.\ these reinforce the 
background magnetic field and might impact on observables 
tied to the early stages of the collisions. They may be 
taken into account in a similar manner as currents
originating from Faraday and Hall effects~\cite{Gursoy:2014aka}.
\acknowledgments
This research was funded by the DFG (Collaborative Research Center CRC-TR 211 ``Strong-interaction matter under
extreme conditions'' - project number 315477589 - TRR 211) and by the Helmholtz Graduate School for Hadron and Ion Research (HGS-HIRe for FAIR). 
The authors are grateful for inspiring discussions
with Massimo D'Elia and Dmitri Kharzeev.
\appendix
\section{Sea-valence decomposition of expectation values}\label{app:sea_valence}
In this appendix, we consider the decomposition of fermionic observables into valence and sea contributions for weak magnetic fields. We follow the discussion of Ref.~\cite{DElia:2011koc} and extend it with one important point. We treat the magnetic field as a continuous variable and consider the Taylor expansion of expectation values in $B$. This is valid in the infinite-volume limit, where the quantum of the magnetic flux approaches zero. The expectation value of an operator $O$ as a function of the magnetic field $B$ for multiple quark flavors can be written as
\begin{align}
\ave{O(B)}_B &= \frac{T}{V\Z(B)}\int \mathcal{D}U e^{-S_g}O(B) \prod_f\det \mathcal{M}_f(B) \\
&= \frac{T}{V\Z(0)}\int \mathcal{D}U e^{-S_g}O(B)\prod_f\det\mathcal{M}_f(0)\frac{\det\mathcal{M}_f(B)}{\det\mathcal{M}_f(0)}\frac{\Z(0)}{\Z(B)} \\
&= \frac{T}{V\Z(0)}\int \mathcal{D}U e^{-S_g} P(B)O(B) \prod_f\det\mathcal{M}_f(0) = \ave{P(B)O(B)}_0\,,
\end{align}
where we defined $P(B) \equiv \prod_f\det\mathcal{M}_f(B)\Z(0)/[\det\mathcal{M}_f(0)\Z(B)]$. The notation $\ave{\cdot}_B$ indicates that the expectation value is taken on configurations at finite $B$. In that fashion, we distinguish between the $B$-dependence in the operator and in the configurations. Each operator in the path integral can be expanded in powers of $B$ as follows
\begin{align}
O(B) = O_0 + O_1 B + O_2 B^2 + \mathcal{O}(B^3)\,, \hspace{1cm} P(B) = P_0 + P_1 B + P_2 B^2 + \mathcal{O}(B^3)\,,
\end{align}
where $P_0=P(0)=1$. Replacing this expansion in the expectation value yields
\begin{align}
\ave{P(B)O(B)}_0 = \sum_{l=0}^\infty\sum_{k=0}^\infty B^{l+k}\ave{P_lO_k}_0\,.
\label{eq:path_int}
\end{align}
Notice that, even if the expectation value $\langle O(B) \rangle_B$ has definite parity, both $O$ and $P$ may contain all powers of the magnetic field under the path integral, as they depend on the particular gauge configuration $U$ on which they are evaluated. This subtle point was ignored in Ref.~\cite{DElia:2011koc}.

Eq.~\eqref{eq:path_int} can be recast as
\begin{align}
\ave{P(B)O(B)}_0 = \ave{P_0O_0}_0 + \sum_{l \neq 0}B^l\ave{P_lO_0}_0 + \sum_{k \neq 0} B^{k}\ave{P_0O_k}_0 + \sum_{l \neq 0, k \neq 0} B^{l+k}\ave{P_lO_k}_0\,.
\end{align}
Defining $O^{\val} \equiv \sum^{\infty}_{k=0} B^k\ave{P_0O_k}_0$ and $O^{\sea} \equiv \sum^{\infty}_{l=0}B^l\ave{P_lO_0}_0$ as the valence and the sea terms, respectively, and noticing that $\ave{P_0O_0}_0 = \ave{O(0)}_0$ the expression above can be written as
\begin{equation}
\ave{O(B)}_B + \ave{O(0)}_0 = O^{\val} + O^{\sea} + \sum_{l \neq 0, k \neq 0} B^{l+k}\ave{P_l O_k}_0\,.
\label{eq:sea_valence}
\end{equation}
Notice that the third term on the right hand side is, since $l\neq0$ and $k\neq0$, at least of $\mathcal{O}(B^2)$. 

In accordance with the definitions~\eqref{eq:val_term} and~\eqref{eq:sea_term} in the main text, the valence and the sea terms for the operator $O$ correspond to the following path integrals
\begin{align}
O^{\val} &= \ave{P(0)O(B)}_0 = \ave{O(B)}_0 = \frac{T}{V\Z(0)}\int \mathcal{D}U e^{-S_g}O(B) \prod_f\det \mathcal{M}_f(0)\,, \\
O^{\sea} &= \ave{P(B)O(0)}_0 = \ave{O(0)}_B = \frac{T}{V\Z(B)}\int \mathcal{D}U e^{-S_g}O(0) \prod_f\det \mathcal{M}_f(B)\,.
\end{align}
The valence term corresponds to measuring the $B$-dependent operator $O(B)$ on $B=0$ configurations, while the sea term corresponds to measuring the $B=0$ operator $O(0)$ on $B$-dependent configurations.

To apply this procedure to the electric current operator, we first make contact with the expectation value of $j_{\mu}$ by defining $O \equiv \Tr(\Gamma_{\mu}\mathcal{M}^{-1})$ inside the path integral. Then, the sea-valence decomposition reads 
\begin{equation}
\ave{j_{\mu}(B)}_B + \ave{j_{\mu}(0)}_0 = j_{\mu}^{\val} + j_{\mu}^{\sea} + \sum_{l \neq 0, k \neq 0} B^{l+k}\ave{P_l \, j_{\mu}^k}_0\,.
\end{equation}
However, the electric current expectation value is an odd function of $B$, hence $\ave{j^{\mu}(0)}_0 = 0$. Moreover, both $j_\mu^{\rm val}$ and $j_\mu^{\rm sea}$ are also odd in $B$ due to parity symmetry, therefore
\begin{equation}
\ave{j_{\mu}(B)}_B = j_{\mu}^{\val} + j_{\mu}^{\sea} + \mathcal{O}(B^3)\,,
\end{equation}
which is Eq.~\eqref{eq:sea_valence_J} of the main text.
Notice that for parity-even observables, like the quark condensate, all three terms in the right hand side of Eq.~\eqref{eq:sea_valence} are $\mathcal{O}(B^2)$ and there is no order in $B$ where the third term may be neglected.
\section{Induced currents in chiral perturbation theory}
\label{app:chiPT}
The low-energy effective theory of QCD is chiral perturbation theory ($\chi$PT). In this appendix, we calculate the induced currents due to the inhomogeneous magnetic field within this framework. In $\chi$PT, the effective degrees of freedom are the pions, among which the charged ones are responsible for the electric current. These are described, to leading order, via scalar QED coupled to the background eletromagnetic field. The theory is thus defined by the Minkowski action
\begin{equation}
    S=\int \dd^4x \left[ \left|D_\mu\Phi\right|^2+m_\pi^2|\Phi|^2 \right],
\end{equation}
where $D_\mu = \partial_\mu-ieA_\mu$ is the covariant derivative in the presence of the background field $A_\mu$ and $m_\pi$ the pion mass.

We turn the argumentation of Sec.~\ref{sec:mag_susc} around: we calculate the momentum-dependent magnetic susceptibility of the charged pions, and use that to build the inhomogenous current with the magnetic field profile used in the main text.
In our choice of gauge ($A_2(x)$ to represent the magnetic field), the magnetic susceptibility can be written as
\begin{equation}
    \widetilde{\chi}(p)=\frac{1}{p^2}\int \dd x \,{\rm e}^{-ixp}\expval{j_y(x)j_y(0)} = \frac{\Pi^{22}(p)}{e^2p^2}\,,
\end{equation}
where $\Pi^{22}$ is the photon polarization tensor of scalar QED and $p$ denotes, just like in Sec.~\ref{sec:mag_susc} of the main text, the momentum in the $x$ direction. The calculations to be made are text-book, and can be found in e.g.\ Chapter 16 of \cite{Schwartz:2014sze}. Two one-loop order diagrams contribute: a momentum-dependent bubble-type, and a momentum-independent tadpole-type. However, when putting all external momentum components, except for $p$, to zero, the final result is considerably simplified, even though both diagrams contribute. In this case the bare polarization tensor in dimensional regularization reads
\begin{align}
    \Pi^{22}(p) = \frac{ie^2}{4\pi^2}\Bigg[&\frac{1}{4-d}\frac{p^2}{6}-\frac{p^2(\gamma_{\rm E}-1)}{12}+\frac{m_\pi^2}{2}\log\frac{m_\pi^2}{4\pi\mu^2}\nonumber\\
    &-\int_0^{1/4}\dd\alpha \frac{m_\pi^2+p^2\alpha}{\sqrt{1-4\alpha}}\log\frac{m_\pi^2+p^2\alpha}{4\pi\mu^2}\Bigg]\,,
\end{align}
where $\gamma_{\rm E}$ is the Euler-Mascheroni constant and $\mu$ the renormalization scale. As well known~\cite{Dunne:2004nc,Endrodi:2013cs}, the prefactor of the divergent term is proportional to the leading-order $\beta$-function coefficient $\beta_1=1/(48\pi^2)$ of scalar QED.

Notice that this singular term leads to a momentum independent divergence for the susceptibility itself, which we can take care of in a similar manner as in the case of full QCD, by a zero-momentum subtraction. This also sets the renormalization scale implicitly, such that
\begin{align}
    \widetilde{\chi}^R(p) &= \frac{\Pi^{22}(p)}{e^2p^2}-\lim_{p\to 0}\frac{\Pi^{22}(p)}{e^2p^2} = -\frac{i}{4\pi^2}\left[\frac{1}{12}-\int_0^{1/4}\dd\alpha\frac{m_\pi^2+p^2\alpha}{p^2\sqrt{1-4\alpha}}\log\frac{m_\pi^2+p^2\alpha}{m_\pi^2}\right]\nonumber\\
    &=-\frac{i}{72\pi^2}\frac{1}{p^3}\left[4\left(3m_\pi^2p+p^3\right)-3\left(4m_\pi^2+p^2\right)^{3/2}{\rm ArcTanh}\frac{p}{\sqrt{4m_\pi^2+p^2}}\right]
\end{align}
is independent of the renormalization scale. We can then calculate the renormalized current profile using~\eqref{eq:chip1} and~\eqref{eq:renorm_curr} as
\begin{equation}
    j^R_y(x) = i \int \dd p\, {\rm e}^{ixp} \,p\,\widetilde{\chi}^R(p)e\widetilde B(p)\,,
\end{equation}
where $\widetilde{B}(p)$ is the Fourier transform of the field profile \eqref{eq:inv_cosh_profile}. 

Calculating the above current using the physical pion mass at weak magnetic fields, one should find reasonable agreement with the continuum extrapolation of our lattice simulation results. Here, we follow a slightly different route instead and incorporate staggered lattice artefacts into the $\chi$PT description. Following~\cite{Huovinen:2009yb,Bali:2020bcn}, we average over the 16, partially degenerate taste copies of pions appearing in staggered lattice simulations, of which only the lightest one has the physical value. The taste splitting of the masses for our action is taken from~\cite{Borsanyi:2010cj}.

\begin{figure}
    \centering
    \includegraphics[width=0.7\linewidth]{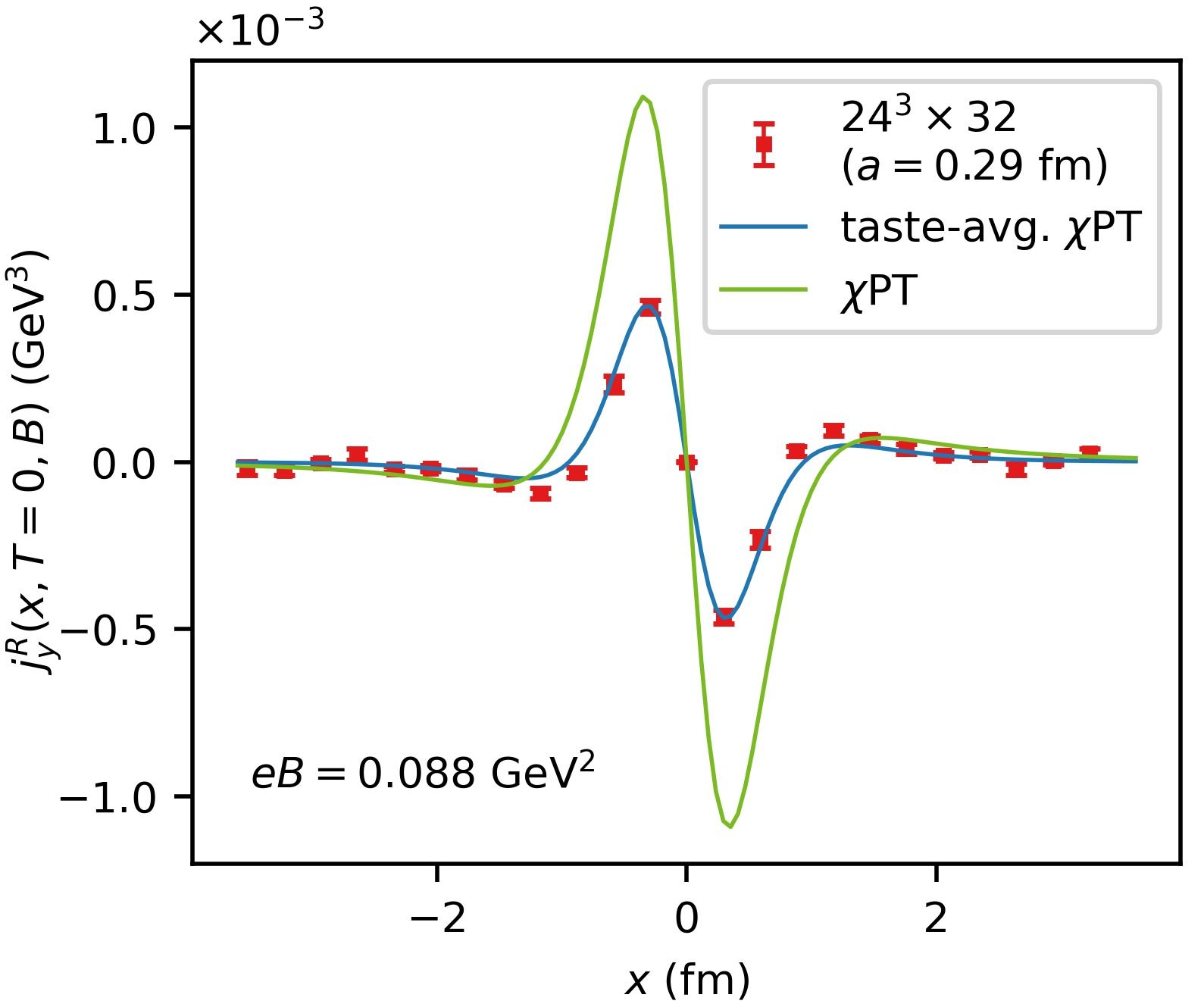}
    \caption{\small Comparison of our lattice data for the renormalized current with chiral perturbation theory estimates. A correct description at finite lattice spacings needs a correction for the breaking of chiral symmetry on the taste level.}
    \label{fig:chiPT_compare}
\end{figure}

Our result for the $\chi$PT prediction of the current is shown in Fig.~\ref{fig:chiPT_compare}, where we plot $j_y^R(x$) for both the continuum setup with the physical pion mass, and for the above described taste-averaged setup. The curves are compared to our lattice data obtained on a $24^3\times32$ lattice at the lowest possible magnetic field, $eB=0.88$ GeV$^2$ with a lattice spacing of $a=0.29$ fm. We observe a remarkable agreement between the lattice data and the taste-averaged $\chi$PT current. In turn, the continuum curve with physical pion mass lies above the lattice results by about a factor of two. This shows that lattice artefacts on this coarse lattice are substantial, but still in a region where a continuum extrapolation is under control.

\bibliographystyle{style}
\bibliography{bibliography.bib}

\end{document}